\documentclass[12pt]{iopart}

\usepackage{iopams}
\usepackage{graphicx}
\usepackage{dcolumn}
\usepackage{bm}
\usepackage{color}
\usepackage{enumerate}

\newtheorem{thm}{Theorem}

\newtheorem{exem}{Example}

\newcommand{\proof}{\noindent {\bf Proof:} \hspace{0.1in}}
\newcommand{\qed}{\hfill\mbox{\raggedright $\Box$}\medskip}

\newcommand*{\num}[1]{\scalebox{0.7}[0.8]{$#1$}}
\newcommand{\tfrac}[2]{\frac{\textstyle{#1}}{\textstyle{#2}}}
\newcommand{\R}{\mathbb{R}}
\newcommand{\metric}{\ensuremath{\mathrm{g}}}
\newcommand{\hubble}{\ensuremath{\mathrm{H}}}
\newcommand{\deviation}{\scalebox{0.7}[0.8]{\ensuremath{\Sigma}}}
\newcommand{\pressure}{\ensuremath{\mathrm{p}}}

\newcommand{\energy}{\ensuremath{\mathrm{\rho}}}
\newcommand{\anisophase}{\ensuremath{\mathrm{\beta}}}
\newcommand{\stressparam}{\ensuremath{\mathrm{\eta}}}
\newcommand{\mixangle}{\ensuremath{\mathrm{\Phi}}}

\newcommand{\mod}{\ensuremath{\mathrm{mod}}}

\begin{document}
\title{The oscillatory anisotropy in the spatially flat cosmological models}

\author{Bruno B. Bizarria\, , \, Gabriel A. Souza Silva}
\address{Astrophysics division - INPE \\
Instituto Nacional de Pesquisas Espaciais, S\~ao Jos\'e dos Campos - SP, Brasil. ZIP:12245-970.}
\eads{\mailto{bruno.bizarria@inpe.br, gabriel.souza@inpe.br}}

\author{Leandro G. Gomes\footnote{Corresponding author}}
\address{Institute of Mathematics and Computation - UNIFEI\\ 
Universidade Federal de Itajub\'a, Av. BPS, 1303,  Itajub\'a-MG, Brasil. ZIP:37500-903}
\eads{\mailto{lggomes@unifei.edu.br}}

\author{William O. Clavijo}
\address{Institute of Mathematics - UFRJ \\ 
Universidade Federal do Rio de Janeiro, Rio de Janeiro - RJ, Brasil. ZIP:21941-909}
\eads{\mailto{woclavijo@ufrj.br}}

%\pacs{}
\date{\today}

%\author{Bruno B. Bizarria\, , \, Gabriel A. S. Silva\\
%\small{Astrophysics division - INPE}\\
%\small{bruno.bizarria@inpe.br, gabriel.souza@inpe.br}\\
%\adress{Instituto Nacional de Pesquisas Espaciais, S\~ao Jos\'e dos Campos - SP, Brasil. ZIP:12245-970.}\\[4mm]
%Leandro G. Gomes\footnote{Corresponding author}\\
%\small{Institute of Mathematics and Computation - UNIFEI}\\
%\small{lggomes@unifei.edu.br}\\
%\tiny{Universidade Federal de Itajub\'a, Av. BPS, 1303,  Itajub\'a-MG, Brasil. ZIP:37500-903 .}\\[4mm]
%William O. Clavijo \\
%\small{Institute of Mathematics - UFRJ}\\
%\small{woclavijo@ufrj.br}\\
%\tiny{Universidade Federal do Rio de Janeiro, Rio de Janeiro - RJ, Brasil. ZIP:21941-909}}

\begin{abstract}
In this manuscript, we investigate the oscillatory behaviour of the anisotropy in the diagonal Bianchi-I spacetimes. Our starting point is a simplification of Einstein's equations using only observable or physical variables. As a consequence, we are able to: (a) Prove general results concerning the existence of oscillations of the anisotropy in the primordial and the late-time universe. For instance, in the expanding scenario, we show that a past weakly mixmaster behaviour (oscillations as we approach the Kasner solutions) might appear even with no violation of the usual energy conditions, while in the future, the pulsation (oscillations around isotropic solutions) seems to be most favored; (b) Determine a large scheme for deriving classes of physically motivated exact solutions, and we give some (including the general barotropic perfect fluid and the magnetic one); (c) Understand the physical conditions for the occurrence of the isotropization or anisotropization during the cosmological evolution; (d) Understand how anisotropy and energy density are converted one into another. In particular, we call attention to the presence of a residue in the energy density in a late-time isotropic universe coming from its past anisotropic behaviour.        
\end{abstract}

\pagebreak

\section{Introduction}

The general perception of the anisotropy in cosmology is that it plays a minor, if not an irrelevant, role. The only exception, if any at all, is during the very early period of the universe, where some kind of chaotic mixmaster behaviour could have taken place \cite{BKL,Mixmaster}. This picture have been assumed due to the success of the standard model of cosmology \cite{ellis_mac_marteens}. From the observational point of view, it has been supported by precise measurements involving the Cosmic Microwave Background radiation (CMB), and more recently, the Baryonic Acoustic Oscillations (BAO), both corresponding to  epochs close to when the radiation was decoupling from matter, that is, at redshift $z \sim 10^3$. At that time, we should expect the anisotropy in the Hubble parameter to be inferior to one part in $10^{11}$ (CMB)\cite{IstSaadeh}, or even less, one part in $10^{15}$ (CMB+BAO)\cite{Tedesco2019}. On the other hand, at small redshift, when we observe the supernovae in the late-time sky, the precision in the observations are far away from that \cite{IstLin,Campanelli,Kalus}, as one part in $10^{2}$ or $10^{3}$, and even some anisotropy detection at this scale might be possible \cite{Schucker,AnstCai,AnstColin}. Notwithstanding, in recent times, doubts have been raised as to the credibility of the standard $\Lambda$CDM model as it is confronted with the observed data \cite{Riess}. Therefore, reviewing its foundations under new perspectives is not mere speculation, but instead, a scientific duty, and it is natural to consider if, as we came towards the construction of a final theory to describe our universe, we could have missed something concerning the anisotropy. 

The first task we are going to tackle is to find some relations connecting the physical conditions on the matter content and the asymptotic behaviour of the solutions. This will be given in theorems \ref{Thm:PastBehaviourGeneral} and \ref{Thm:FutureBehaviourOscillatory}. The general context underlining those results can be seen, for instance, as we investigate the asymptotic past of the cosmological models. It has been done in two fronts: one is concerned with the existence and robustness of the primeval singularity, while the other, on the behaviour of the spacetime as this epoch is approached. It is fair to say that, in the classical level, the first of them is better understood and lays on a strong mathematical foundation \cite{haw_ellis,Wald}. On the other hand, the second one is mainly guided by the BKL picture \cite{BKL}, which, roughly speaking, says that the dominant part of the dynamics, as we come close to the Big Bang, is just as in the spatially homogeneous models. Although this scheme have not already achieved the accuracy level of the singularity theorems, since a comprehensive and rigorous mathematical description seems to be still missing \cite{Ring}, the physical argument is compelling and many analytical and numerical considerations favors it (See \cite{BGIMW,Berger,BelinskiLivro} and the references therein). This means that the behaviour of the spacetime in such early ages is well described by the class of homogeneous cosmological models. Hence, understanding their asymptotic dynamics in different situations is the first step towards the construction of a consistent picture of the universe's possible histories. Despite the fact that this aspect in the Bianchi-I dynamics have been quite studied in the literature (see \cite{calogero,wainwright} and the references therein), we are going to unveil some of their interesting features that have been hidden so far, as for instance, some general  conditions sufficient for the appearance of the weakly mixmaster behaviour. We touch on this subject all along the section \ref{Sec:AnisotropyTheorems}. In the section \ref{Sec:ExactSolution}, we examine the new theorems \ref{Thm:PastBehaviourGeneral} and \ref{Thm:FutureBehaviourOscillatory} on the light of some specific equations of state. They provide not only good examples to understand the general picture, but also new classes of physically motivated exact solutions. In particular, we obtain a large scheme for generating new exact solutions from a two dimensional system of ordinary differential equations.   

The isotropization, that is, the property of the universe to get more isotropic as the time passes by, is tacitly taken for granted in the standard model of cosmology, even as the general theorems favor the opposite, the anisotropization. In fact, it is shown that for a homogeneous initial condition, the isotropization will occur only in a set of measure zero in the space of the cosmological parameters \cite{Hawking}. Hence, investigating the physical conditions under which each process occur in the Bianchi-I spacetimes is demanding. We do it in the section \ref{Sec:IsotropizationAnisotropization}.   

There is an important parameter that is missing in the perfect fluid case: the anisotropy phase $\anisophase$ (see formula (\ref{Eq:DiagAnisoStress})). It appears when the the anisotropy in the matter components is taken into account, and it is a measurement on how it is coupled to its counterpart in the Hubble parameter. In this case, the anisotropies start to bounce around the different directions. This motion can be converted into energy, and vice-versa, depending on the signal of $\cos\anisophase$. In the section \ref{Sec:EnergyAnisotropyExchange}, we investigate how it can happen and some of its consequences. In particular, we call attention to the presence of a residue in the energy density in a late-time isotropic universe coming from its past anisotropic behaviour, and this is nothing like the usual $a^{-6}$ term often considered as the typical behaviour of the anisotropy effective ``energy density'' \cite{Tedesco2019}.        

We finish our manuscript with a self-criticism in the section \ref{Sec:CriticismOscillations}. There, we analyse the effectiveness of the definition of oscillation and mixmaster behaviour, both given in the section \ref{Sec:AnisotropyTheorems}. As we are going to see, they both are too evasive to be easily tamed by a precise mathematical definition.

\section{The Einstein's equations and the polar representation of the Kasner disk}

Our starting point is to restrict the dynamical variables to those with a direct physical or observational significance. So, instead of working with the ``Mini Super Space'' \cite{Berger,Misner} or the Calogero and Heinzle's parameters \cite{calogero}, we deal with $\hubble_1$, $\hubble_2$ and $\hubble_3$, the Hubble factors measured along the principal orthogonal directions. They are well described by the isotropic Hubble parameter $\hubble:= (\hubble_1+\hubble_2+\hubble_3)/3$, the anisotropy magnitude $\deviation$, which is also referred to as ``shear parameter" \cite{wainwright} or ``Cosmic Shear" \cite{Tedesco},  and  the anisotropy polar angle $\alpha$:
\begin{equation}\label{Eq:DeterminationDeviationPhase}
\deviation = 
\sqrt{\frac{1}{6}\, \sum_{k=1}^3\,\left(\frac{\hubble_k-\hubble}{\hubble}\right)^2} 
\quad \textrm{and} \quad
\sin \alpha=\frac{\hubble_1-\hubble}{2\, \deviation \, \hubble} \, .
\end{equation}
This is equivalent to write (compare with $\sigma_k=\hubble_k - \hubble$ in the reference \cite{BGK}), 
\begin{equation}\label{Eq:HubbleFactorsDirections}
\deviation \,  \sin\left(\alpha + \varphi_k\right) = \frac{1}{2}\, \frac{\hubble_k-\hubble}{\hubble}
\qquad (\, \varphi_{k+1} =2k\pi/3\,) \, .
\end{equation}
The metric is represented as
\begin{equation}\label{Eq:GeneralMetric}
\metric = -\,dt^2 \,+
\ell_1^2 \, dx^2 \,+\, \ell_2^2 \, dy^2 \,+ \, \ell_3^2 \, dz^2 
\end{equation}
with
\begin{equation}\label{Eq:ScaleFactors}
\ell_k = \ell_k(t_0) \, e^{b_k(t)/2} \quad , \quad   b_{k} (t):=  2\,\int_{t_0}^{t} \, \hubble \, 
\left(1+2\, \deviation \,  \sin\left(\alpha + \varphi_k\right)\, \right) \, dt\, .
\end{equation}
As usual, we define the isotropic scale factor as $a=(\ell_1\ell_2\ell_3)^{1/3}$, that is, $\hubble=\dot{a}/a$.

In a similar way, following a scheme analogous to the one presented in \cite{BGK}, the matter is represented by the energy density $\energy$, the relativistic pressure $\pressure$ and the diagonal and traceless anisotropic stress tensor, $(\pi_{ij}) = \textbf{diag} \{\pi_1,\pi_2,\pi_3\, \}$, which we describe by the stress relative magnitude $\stressparam$ and the anisotropy phase $\anisophase$, determined by 
\begin{equation}\label{Eq:DiagAnisoStress}
\stressparam \, \sin(\alpha+\anisophase+\varphi_k) = \frac{\pi_k}{\energy}  \, .    
\end{equation}
Under these variables, the first two Einstein's equations are represented by the generalized Friedmann one, 
\begin{equation}\label{Eq:Friedmann}
3\, \hubble^2\, \left(\, 1 - \deviation^2 \, \right) \, = \, \energy  \, ,
\end{equation}
and the conservation equation,\footnote{Unless specified, $\gamma$ is not constant.} 
\begin{equation}\label{Eq:Conservacao}
\dot{\energy} = -\, 3\, \hubble \, \energy \, \left( \gamma  + \deviation \, \stressparam \cos{\anisophase} \right)
\qquad \left( \, p=(\gamma-1)\energy \, \right)\, .
\end{equation}
As we assume $\dot{\energy}\ne 0$, we can take the time-like coordinate $s$, which measures the order of magnitude of $\energy$,  
\begin{equation}\label{Eq:DefinitionVariableS}
s = \ln\left(\,\frac{\energy_0}{\energy}\,\right) \, .
\end{equation}
Hence, the last equations are 
\begin{equation}\label{Eq:AnisotropyMagnitudeEquation}
\frac{d\deviation}{ds}=
- \, \frac{(1-\deviation^{2})}{2}\left(\frac{(2-\gamma)\,\deviation - \stressparam \cos{\beta}}{\gamma  + \deviation \, \stressparam \cos{\anisophase}}\right)
\end{equation}
and 
\begin{equation}\label{Eq:AlphaEquation}
\frac{d\alpha}{ds}=  \frac{(1-\deviation^{2})}{2\, \deviation}\left(\frac{\stressparam\sin{\beta}}{\gamma + \deviation \,\stressparam\cos{\anisophase}}\right) \, .
\end{equation}
These two equations represent the projection of the dynamical system in the Kasner disk \cite{calogero}, that is, the set $\deviation_1^2+\deviation_2^2+\deviation_3^2 \le 6$ in the plane $\deviation_1+\deviation_2+\deviation_3=0$, where $\deviation_k=(\hubble_k-\hubble)/\hubble=2\, \deviation \,  \sin\left(\alpha + \varphi_k\right)$. In fact, our representation is given in terms of the ``radial'' coordinate $\deviation$ and the polar angle $\alpha$. Hence, we have a polar representation of the Kasner disk, where $0\le \deviation \le 1$.  Note that the Kasner circle $\deviation=1$ is composed by the vacuum Kasner solutions \cite{ellis_mac_marteens,Kasner} , while the center of the disk represents the isotropic flat Robertson-Walker (RW) spacetimes ($\deviation=0$).\footnote{In fact, there are solutions with $\deviation=1$, non-vanishing pressure and zero energy density. They are avoided if, for instance, we assume $\gamma$ and $\stressparam$ bounded.} 
\begin{figure}
\begin{center}
\includegraphics[width=0.35\textwidth]{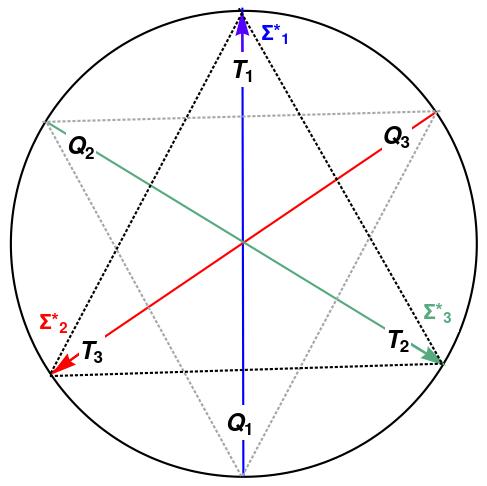} 
\hspace{2cm}
\includegraphics[width=0.35\textwidth]{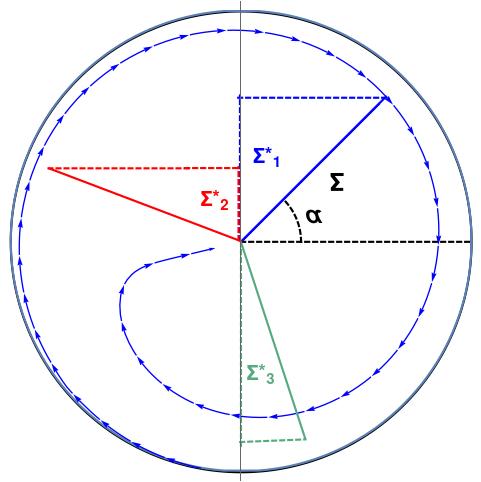}
\caption{The ``polar" Kasner disc and the Taub points with $\Sigma^{*}_{k}=\frac{\Sigma_{k}}{2}$.}
\end{center}
\label{Fig:KasnerVsPolar1}
\end{figure}

The simplifying formulas presented so far allow us to investigate the general conditions on the matter content for which there is an asymptotic oscillatory behaviour, that is, the polar angular frequency, given by   
\begin{equation}\label{Eq:AngularFrequency}
\omega := \dot{\alpha} =  \frac{3}{2}\, \stressparam\, \hubble\,(1-\deviation^{2}) 
 \, \frac{\sin \anisophase}{\deviation} \, ,
\end{equation}
does not converge to zero asymptotically. When a solution tends to show this behavior arbitrarily close to the Kasner circle, we say that it is weakly mixmaster: an asymptotic regime characterized by alternating Kasner periods. This notion was first introduced in the context of the Bianchi IX models, in the late sixty's \cite{Mixmaster,BKL}, and latter found in the Bianchi-I spacetimes \cite{LeBlanc,calogero}. As we will show through the theorems \ref{Thm:PastBehaviourGeneral} and \ref{Thm:FutureBehaviourOscillatory}, there are general sufficient conditions to ensure the occurrence of this phenomena.

\section{The anisotropy oscillations}\label{Sec:AnisotropyTheorems}

In this section, we shall focus our attention on the analysis of the anisotropic dynamics, or, to be more specific, that part described by the parameters $\deviation$ and $\alpha$ in the polar representation of the Kasner disk. The center of the disk is the isotropic point ($\deviation=0$), where the spacetime turns (flat) Robertson-Walker, while any point in the exterior Kasner circle ($\deviation=1$) is a Kasner vacuum solution. We will examine the general conditions under which the matter content impinges an asymptotic oscillatory behavior to the expanding universe, as of a weakly mixmaster type (alternating Kasner epochs as we approach the singularity), as of a pulsating type (oscillations as it approachs the isotropic point) or even as something in between these two regimes. 

Here we shall concentrate in the mathematical aspect of the theory: put the definitions precise and derive the useful theorems correctly. We assume each solution of the field equations to be given in its maximal interval of definition $(t_{-},t_{+})\subset\R$ with polar angular frequency $\omega(t)$. \footnote{ By field equations we mean the Einstein's plus the equations coming from the matter content (algebraic phenomenological equations, scalar fields, electromagnetic fields,...)} We define:    
\begin{enumerate}[(a)]
\item A solution is isotropic in the future (past) if $\deviation \to 0$ as $t \to t_{+}$ ($t \to t_{-}$). 
\item A solution is Kasner in the past (future) if $\deviation \to 1$ as $t \to t_{-}$ ($t \to t_{+}$).  
\item A solution is oscillatory in the past (future) if there exists $\Omega > 0$ and a strictly decreasing (increasing) sequence $\{ t_n \}$ converging to $t_{-}$ ($t_{+}$) such that $|\omega(t_n)|\ge \Omega$ for every $n$.
\item A solution is weakly mixmaster in the past (future) if there exists $\Omega > 0$ and a strictly decreasing (increasing) sequence $\{ t_n \}$ converging to $t_{-}$ ($t_{+}$) such that $|\omega(t_n)|\ge \Omega$  for every $n$ and $\deviation(t_n) \to 1$. 
\item A solution is pulsating in the future (past) if it is both isotropic and oscillatory in the future (past).
\end{enumerate}

We left to the section \ref{Sec:CriticismOscillations} our further comments relating to the interpretation and criticism on the definitions of the mixmaster and the oscillatory behaviours. The reader who fells uneasy at this point, should consider reading that section first.  

Note that if a solution is not oscillating in the past, then for every sequence $t_n$ converging to $t_{-}$, we have that $\omega(t_n)$ converges to zero. This means that $\omega(t)\to 0$ as $t\to t_{-}$. On the other hand, the oscillatory behavior will be present when $\omega\to \Omega \ne 0$ or this limit is not well defined. The same is true in the future, replacing $t_{-}$ by $t_{+}$.

The theorems in this section are formulated in the context of the cosmological expansion, $\hubble >0$, but can be straightforwardly adapted to the case of contraction, $\hubble <0$. The first of them deals with the asymptotic behaviour in the past (``$\alpha$-limit"), while the other, in the future (``$\omega$-limit").
\begin{thm}\label{Thm:PastBehaviourGeneral}
Consider a solution of the Einstein's equations in its maximal interval of definition $(t_{-},t_{+})$. Let $T > t_{-}$ be a real number with the property that, along the interval $(t_{-},T)$, we have $\hubble > 0$ , $0 < \deviation <1$ and $\gamma$ and $\stressparam$ are bounded.
\begin{enumerate}[(i)]
\item Assume the solution to be Kasner in the past with the constant $\xi_{+}$ satisfying
\begin{equation}\label{Eq:PastBehaviourGeneralMais}
 \gamma + \stressparam\, \cos\anisophase  \le \xi_{+} < 1 
\end{equation}
along the interval $(t_{-},T)$. If $a\to 0$ as $t\to t_{-}$, then the solution is non-oscillatory in the past.
\item Assume the solution to be Kasner in the past with the constant $\xi_{-}$ satisfying
\begin{equation}\label{Eq:PastBehaviourGeneralMenos}
1 <  \xi_{-} \le \gamma + \stressparam\, \cos\anisophase  
\end{equation}
along the interval $(t_{-},T)$. If $\stressparam\,\sin \anisophase$ does not converge to zero as $t\to t_{-}$, then the solution is weakly mixmaster in the past. In the regime $t \to t_{-}$, if $\stressparam\,\sin \anisophase$ is bounded away from zero and $a\to 0$, then $|\omega| \to \infty$.
\item Assume the solution is not Kasner in the past, such that $\deviation$ is bounded away from $1$ in the interval $(t_{-},T)$. If $\gamma \ge 0$ and the limit for $\stressparam\,\sin \anisophase$ as $t\to t_{-}$, if it exists, is not zero, then the solution is oscillatory in the past. In particular, if it is isotropic in the past with $\stressparam\,\sin \anisophase$ bounded away from zero, then $|\omega| \to \infty$ as $t\to t_{-}$.
\end{enumerate}
\end{thm}
\proof 
Let $u(t)$ satisfy $\dot{u}=-3\, u\, \hubble\, (\gamma + \stressparam\, \cos\anisophase)$ and take the constants $\xi_{\pm}$ such that, in the interval $(t_{-},T)$,
\begin{equation}
\xi_{-} \le \gamma + \stressparam\, \cos\anisophase \le \xi_{+}\, . 
\end{equation}
Since $\hubble >0$, we have
\begin{equation}
 3\, \hubble\, \xi_{-} \le -\, \frac{\dot{u}}{u}  \le 3\, \hubble\, \xi_{+}\, .
\end{equation}
As we integrate it from $t$ to $T$, $t_{-}<t<T$, we get 
\begin{equation}
\left(\frac{a(T)}{a(t)}\right)^{3\, \xi_{-}} \le \frac{u(t)}{u(T)}  \le \left(\frac{a(T)}{a(t)}\right)^{3\, \xi_{+}} \, .
\end{equation}
Now, define $z(t)$ by $\energy = u\, a^3 \, \hubble \, e^{z}$. Applying the Einstein's equations (\ref{Eq:Friedmann})-(\ref{Eq:AlphaEquation}), we get
\begin{equation}\label{Eq:DeviationDot}
\dot{\deviation} = - \, \frac{3}{2}\, \hubble \, (1-\deviation^2)\,\left(\,(2-\gamma)\,\deviation - \stressparam\, \cos \anisophase\,\right) 
\end{equation}
\begin{equation}\label{Eq:HubbleDot}
\dot{\hubble}= - \, \frac{3}{2}\, \hubble^2\, \left((1-\deviation^2)\, \gamma + 2 \deviation^2\right) 
\end{equation}
and 
\begin{equation}
\dot{z} = 
-\, \frac{3}{2}\, \hubble \, \, (1-\deviation)\,\left(\,(2-\gamma)\,(1+\deviation) - 2\, \stressparam\, \cos \anisophase\,\right) \, . 
\end{equation}
Let $\epsilon_{-}$ and $\epsilon_{+}$ be constants such that
\begin{equation}\label{Eq:PastBehaviourGeneral2}
2\, \epsilon_{-} \le 
(1-\deviation)\,\left(\,(2-\gamma)\,(1+\deviation) - 2\, \stressparam\, \cos \anisophase\,\right)  
\le 2\, \epsilon_{+} \, .
\end{equation}
If we proceed just as we did to $u(t)$, we arrive at  
\begin{equation}
\left(\frac{a(T)}{a(t)}\right)^{3\, \epsilon_{-}} \le e^{z(t)-z(T)}  \le \left(\frac{a(T)}{a(t)}\right)^{3\, \epsilon_{+}} \, .
\end{equation}
We put them together back in $\energy$, thus arriving in the inequalities
\begin{equation}\label{Eq:PastBehaviourGeneral3}
A_0\, \left|\,\frac{\stressparam\, \sin\anisophase}{\deviation}\,\right|\, a^{3(1-\xi_{-}-\epsilon_{-})} 
\le  \frac{\energy}{\hubble} \, \left|\,\frac{\stressparam\, \sin\anisophase}{2\, \deviation}\,\right|\le
B_0\, \left|\,\frac{\stressparam\, \sin\anisophase}{\deviation}\,\right|\, a^{3(1-\xi_{+}-\epsilon_{+})} \, .    
\end{equation}
for some $0<A_0<B_0$. But according to the definiton (\ref{Eq:AngularFrequency}) and the equation (\ref{Eq:Friedmann}), at the center of these inequalities we have $|\omega|$, proving, for $t_{-}<t<T$, 
\begin{equation}
A_0\, \left|\,\frac{\stressparam\, \sin\anisophase}{\deviation}\,\right|\, a^{3(1-\xi_{-}-\epsilon_{-})} 
\le |\omega| \le
B_0\, \left|\,\frac{\stressparam\, \sin\anisophase}{\deviation}\,\right|\, a^{3(1-\xi_{+}-\epsilon_{+})} \, .    
\end{equation}

If the solution Kasner is in the past, we can make $\epsilon_{\pm}$ arbitrarily small, just by taking $T$ properly such that $(1-\deviation)$ becomes small. In this case, they will not affect neither the sign of $1-\xi_{-}$ nor $1-\xi_{+}$. Therefore, as $t\to t_{-}$, if $\xi_{+}<1$ and $a\to 0$ then $\omega \to 0$, thus proving the first assumption in the theorem. The second one follows in a similar way, as we note that in the case $1<\xi_{-}$, the condition of $\stressparam\sin\anisophase$ not converging to $0$ implies the existence of a sequence  $t_n\to t_{-}$
\[
\Omega_n =A_0\, \left|\,\frac{\stressparam(t_n)\, \sin\anisophase(t_n)}{\deviation(t_n)\, a(t_n)^{3(\xi_{-}+\epsilon_{-}-1)}}\,\right| \le |\omega(t_n)|
\]
bounded away from zero (note that $\hubble >0$ implies $a(t)$ bounded in $(t_{-},T)$). This proves that the solution is weakly mixmaster in the past. Moreover, if $\stressparam(t)\, \sin\anisophase(t)$ is bounded away from zero, the formula above holds for every $t<T$, implying the last statement of the second part of the theorem.

On the other hand, if there is a constant $\deviation_{-}$ with $0\le \deviation\le \deviation_{-} < 1$ along $(t_{-},T)$, and taking the formula (\ref{Eq:AngularFrequency}) into account, the condition of $\stressparam\sin\anisophase$ not converging to $0$ implies the existence of a sequence $t_n \to t_{-}$ with  
\[
\omega(t_n) = \frac{3}{2}\, \hubble(t_n) \,(1-\deviation(t_n)^{2}) 
 \, \frac{\stressparam(t_n)\, \sin \anisophase(t_n)}{\deviation(t_n)} 
\]
bounded away from zero. Here we have used that $\gamma \ge 0$ implies $\dot{\hubble} \le 0$, according to the equation (\ref{Eq:HubbleDot}). In other words, $\hubble \ge \hubble(T)>0$ in $(t_{-},T)$. Note also that if $\stressparam\sin\anisophase$ is also bounded away from zero and $\deviation \to 0$, then $|\omega| \to \infty$ as $t\to t_{-}$. This completes the proof of the theorem.
\qed

On the other hand, looking to the future asymptotic behaviour, we have the general theorem: 
\begin{thm}\label{Thm:FutureBehaviourOscillatory}
Consider a solution of the Einstein's equations in its maximal interval of definition $(t_{-},t_{+})$. Let $T < t_{+}$ be a real number with the property that, along the interval $(T,t_{+})$, we have $\hubble > 0$ , $0 < \deviation <1$ and $\gamma$ and $\stressparam$ bounded.
\begin{enumerate}[(i)]
\item Assume the solution to be isotropic in the future with $\stressparam$ bounded away from zero. It is pulsating in the future if $\cos\anisophase \le \cos\anisophase_{-} < 0$ along $(T,t_{+})$ for some fixed $\anisophase_{-}$ and $\sin\anisophase$ does not converge to $0$ in the $t_{+}$ limit.    
\item Assume the solution to be isotropic in the future with $\stressparam$ bounded away from zero and $t_{+}=\infty$. It is non-oscillatory in the future if $\cos\anisophase \ge \cos\anisophase_+>0$ along $(T,\infty)$ for some fixed $\anisophase_{+}$.
\item Suppose that $\deviation$ is bounded away from $0$ with $\gamma \ge 0$ for every $t>T$, where we have assumed $t_{+}=\infty$. The solution is not oscillatory in the future.
\end{enumerate}
\end{thm}
\proof 
First, let $\deviation \to 0$ as $t\to t_{+}$. As we define $v=\deviation/\hubble$, we get from (\ref{Eq:DeviationDot}) and (\ref{Eq:HubbleDot})
\begin{equation}\label{Eq:FutureBehaviourOscillatoryProof1}
\omega =  \frac{3}{2}\,  (1-\deviation^{2}) 
 \, \frac{\stressparam\sin \anisophase}{v} 
 \quad \textrm{with}\quad
 \dot{v} = \frac{3}{2}\, \stressparam \cos\anisophase + \deviation\, A \, ,
\end{equation}
where we have defined the bounded function $A$ as
\begin{equation}
A(t)= 3(\gamma-1)-\frac{3}{2}  (\eta \cos\anisophase)\, \deviation +3(2-\gamma)\, \deviation^{2}    \, .    
\end{equation} 
Assume $\stressparam \ge \stressparam_{min} >0$ for some constant $\stressparam_{min}$. If  $\cos\anisophase \le \cos\anisophase_{-} < 0$, then $v(t)$ is decreasing close to the isotropic point. As $\sin\anisophase$ does not tend to $0$ in the $t_+$ limit, there is a sequence $\{t_n\}$ such that $\omega(t_n)$ is bounded away from zero, thus proving the first statement. On the other hand, if $\cos\anisophase \ge \cos\anisophase_{+} > 0$, from the mean value theorem, we conclude that $v(t)-v(T) \ge B(t^*)\, (t-T)$ whenever $t>T$ for some $T<t^*<t$, where $B(t)= \stressparam_{min}\, \cos\anisophase_{+} + \deviation(t)\, A(t)$. Since $A$ is bounded, we can set $T$ such that $B(t)>\epsilon>0$ for every $t>T$, $\epsilon$ constant. Therefore, as $t\to \infty$, we get $\omega \to 0$, proving the second statement.

In order to prove the last statement, let $0< \epsilon < \deviation$ and $\gamma \ge 0$ along the interval $(T,t_{+})$, $\epsilon$ constant. Hence, there is a constant $\epsilon_{-}$ such that, using the equation (\ref{Eq:HubbleDot}), 
\[
0 < \epsilon_{-} \le -\, \frac{\dot{\hubble}\,}{\hubble^2}=\frac{3}{2}\, ((1-\deviation^2)\, \gamma + 2 \deviation^2) \, .
\]
Defining $c=1/\hubble(T)$ and integrating from $T$ to $t$, for $t>T$, we get 
\[
0 < \hubble(t) \le \frac{1}{\epsilon_{-}\, (t-T) +c }
\quad \Rightarrow \quad
\lim_{t\to \infty}\hubble(t)=0\, .
\]
Since $\deviation(t)>\epsilon$, we also conclude that $\omega \to 0$ as $t\to \infty$.  
\qed

There are many qualitative studies covering the asymptotic behaviour of the Bianchi I dynamics for specific conditions, as the LRS case \cite{calogero2} ($\alpha=\pm \pi/6, \pm \pi/2,  \pm 5\pi/6$ and $\sin\anisophase=0$), the pure magnectic field plus a $\gamma$-law matter source \cite{LeBlanc} and the magnetic field with a viscous fluid  \cite{Kohli}. None of them have concentrated on the specific criteria for the occurrence of oscillations, so that our analysis, besides being more general in the arbitrariness of the matter content, is complementary to theirs. A good example is the work in the reference \cite{calogero}, which includes some cases as the pure magnetic field and collisionless or elastic matter. It relies on the asymptotic approach to the boundaries of an abstract space formed by the extended ``solutions'', where the ratio of one of scale factors have become arbitrarily greater than the others, as if we could set, loosely speaking, ``$\ell_i/\ell_k = \infty$''. The reader will find a major resemblance of their work in our subsection \ref{Sec:AlphaPlusBetaConstant}. This follows from the assumptions they have assumed: $\gamma$ is constant during the dynamics, while in the asymptotic regime, $\pi_i/\energy$ is also constant, that is, $\stressparam$ and $\alpha+\anisophase$ are asymptotically constant. On the other hand, in our theorems, we are demanding no more than the existence of bounds for $\gamma$ and $\stressparam$, whatever they could be. Furthermore, our results come as complementary to theirs, so that they should be seen as adding to, rather than generalizing, the previous results on the Bianchi-I dynamical systems.    

There is a subtle aspect of the theorem \ref{Thm:PastBehaviourGeneral} that we shall clarify to use it latter. Let us assume that 
\begin{equation}\label{Eq:ParameterQualitative}
\xi_1 = \lim_{t\to t_{-}}\, \left(\gamma+\stressparam\,\cos\anisophase\right)
\end{equation}
is well defined for a solution which is Kasner in the past. As we take $\hubble$ from the equation (\ref{Eq:Conservacao}) and integrate it, we obtain  
\begin{equation}
\ln \left( \frac{a(t)}{a_0}\right) = -\, \frac{1}{3}\, \int_{\energy_0}^{\energy(t)} \frac{1}{\gamma+ \deviation\,\stressparam\,\cos\anisophase}\, \frac{d\energy}{\energy} \, .
\end{equation}
If we take an instant $t_0$ when $\deviation(t_0)$ is arbitrarily close to $1$, we can easily conclude that
\begin{equation}\label{Eq:ScaleFactorKasnerCircle}
a \to a_0\, \left(\frac{\energy_0}{\energy}\right)^\frac{1}{3\,\xi_1} \quad \textrm{if} \quad t \to t_{-} \, ,
\end{equation}
where we have assumed $\xi_1\ne 0$. Therefore, the Kasner asymptotic behaviour of the scale factor and the energy density are connected by the value of $\xi_1$. Since our solutions have physical meaning only when $\energy \ge 0$ and $\dot{\energy}\, \hubble \le 0$, we should expect $\xi_1 >0$, in general. In other words, $a \to 0$ is ``physically equivalent" to $\energy \to \infty$ in the Kasner period.

\section{Some phenomenological models, their asymptotic oscillations and exact solutions}
\label{Sec:ExactSolution}

In this section we specify the matter source in the form of some phenomenological equations of state. In order to determine the full dynamics containing the Einstein's equations, which are $4$ in number for $7$ free variables, we need at least $3$ of them, which will come in the general algebraic form
\begin{equation}\label{Eq:EquationStateOriginal}
\pressure = \pressure (\energy, \hubble_1, \hubble_2, \hubble_3) 
\,\,  \textrm{and} \,\, 
\pi_{ij} = \pi_{ij} (\energy, \hubble_1, \hubble_2, \hubble_3) \, .
\end{equation}
These are the most general phenomenological equations relating the pressure to the energy density and the observational Hubble parameters.  According to the relations (\ref{Eq:HubbleFactorsDirections}), the constraint (\ref{Eq:Friedmann}) and the definition of the variable $s$ in the equation (\ref{Eq:DefinitionVariableS}), this is equivalent of assuming the algebraic equations of state in the form 
\begin{equation}\label{Eq:EquationState}
\gamma = \gamma (s,\deviation, \alpha) \, , \,
\stressparam = \stressparam (s,\deviation, \alpha) 
\,\,  \textrm{and} \,\, 
\anisophase = \anisophase (s,\deviation, \alpha) \, .
\end{equation}
Their choice will be physically or conceptually motivated. Note that variables $\alpha$ and $\anisophase$ are defined mod $2\pi$, so that any relation involving them must respect this condition. Since they can be view as coordinates for the circle $S^1$, the equations of state (\ref{Eq:EquationState}) define a map among open sets of the manifold $\mathbb{R}^2\times S^1$.

From the mathematical viewpoint, the dynamics now is reduced to the two dimensional system in the Kasner disk, given by the equations (\ref{Eq:AnisotropyMagnitudeEquation}) , (\ref{Eq:AlphaEquation}) and (\ref{Eq:EquationState}). The conditions on existence, uniqueness and continuous dependence on the initial values are easily obtained from the classical results of the theory of Ordinary Differential Equations \cite{Arnold}. Hence, the polar Kasner disk variables define the whole dynamics! In fact, if $(\deviation(s),\alpha(s))$ is a solution, after changing ``$s$'' by ``$\ln(\energy_0/\energy)$'', just as in the equation (\ref{Eq:DefinitionVariableS}), we recover its canonical time variable $t$ by integrating the conservation equation (\ref{Eq:Conservacao}), that is, \footnote{Note that each term in the integral is a known function of $\energy$, for since $(\deviation(\energy),\alpha(\energy))$ is given, we also have $\gamma(\energy)$, $\stressparam(\energy)$ and $\anisophase(\energy)$ from the equations (\ref{Eq:EquationState}).}
\begin{equation}\label{Eq:TimeDependenceEnergy}
 t = t_0 \pm \, \int_{\energy_0}^{\energy(t)} \, \sqrt{\frac{1-\deviation^2}{3\, \energy^3}\,}\, \frac{d\energy}{\gamma  + \deviation \, \stressparam \cos{\anisophase}} \, ,
\end{equation}
where $\pm$ is the signal of $-\,\hubble$.  This also represents the solution $\rho(t)$ in an implicit form. The solution of the Einstein's equations are completed as we set the metric (\ref{Eq:GeneralMetric}) with 
\begin{equation}\label{Eq:ScaleFactorsExactSolutions}
b_{k+1} (t):=  - \frac{2}{3}\, \int_{\energy_0}^{\energy(t)} \, \left(\frac{1+ 2\, \deviation\, \sin\left(\alpha + \num{2k\pi/3} \right)  }{\gamma  + \deviation \, \stressparam \cos{\anisophase}}\right) \, \frac{d\energy}{\energy}\, .
\end{equation}
This is a large scheme of deriving exact solutions for the Bianchi-I spacetimes, which involves only physical and observational parameters.

\subsection{The general barotropic perfect fluid solution}\label{Sec:PerfectFluid}

The first, and the simplest, class of examples we deal with is the general barotropic perfect fluid: 
\begin{equation}
\pressure=\pressure(\energy) \quad \textrm{and} \quad \pi_{ij}=0  \quad (\stressparam=0)\, .    
\end{equation}
They form the typical solution with null angular frequency $\omega=\dot{\alpha}=0$, which is readily verified from the equation (\ref{Eq:AlphaEquation}). Hence, they are represented by radial straight lines in the polar Kasner disk. The equation (\ref{Eq:TimeDependenceEnergy}) turns out to be
\begin{equation}\label{Eq:PerfectFluidGeneralSolution}
t = t_0 \pm \int_{\energy_0}^{\energy(t)} 
\, \sqrt{\frac{1- \deviation(\energy)^2}{3 \, \energy}} \, \frac{d \energy}{\energy + \pressure(\energy)} \, ,
\end{equation}
where
\begin{equation}\label{Eq:DeviationEnergyRepresentation}
\deviation (\energy)=  \frac{ \deviation_0}{\sqrt{\deviation_0^2 + (1-\deviation_0^2)\, e^{-G(\energy)}\,}} \qquad G(\energy)= \int_{\energy_0}^{\energy} \, \frac{\energy'-\pressure(\energy')}{\energy'+\pressure(\energy')} \, \frac{d\energy'}{\energy'} \, .
\end{equation}
The metric is determined by the formula (\ref{Eq:ScaleFactorsExactSolutions}) with $\alpha=\alpha_0$, $\stressparam=0$, $\gamma(\energy)=1+p(\energy)/\energy$ and $\deviation(\energy)$ in (\ref{Eq:DeviationEnergyRepresentation}). These formulas simplify the general scheme for obtaining the exact solutions in the barotropic case, as it is presented in \cite{BGK,ExactSolutions}. 

\begin{exem}\label{Exe:PerfectFluidConstantGamma}
Assume \footnote{The case $\gamma=2$ is trivially integrated, for $\deviation(t)=\deviation_0$.} $\gamma\ne 2$ constant. Integrating the equation (\ref{Eq:AnisotropyMagnitudeEquation}) in the variable $s$ and then changing to $\energy$ through the relation (\ref{Eq:DefinitionVariableS}), we get:
\begin{equation}\label{Eq:DeviationEnergyRepresentationGammaConstant}
\frac{\energy}{\energy_0} = 
\left(\frac{\deviation^2\, (1-\deviation_0^2)}{\deviation_0^2\, (1-\deviation^2)}
\right)^{\kappa_1}
\quad , \, \kappa_1= \frac{\gamma}{(2-\gamma)} \, .
\end{equation}
Note that, for $0 < \gamma < 2$, as we approach the Kasner epoch ($\deviation \to 1$) the energy density becomes arbitrarily large, while in the isotropic era ($\deviation \to 0$) it tends to zero. This is the global picture commonly associated to the Bianchi-I context \cite{wainwright}. Returning to the cosmic time $t$, the relation (\ref{Eq:TimeDependenceEnergy}) is simpler expressed in terms of $\deviation$, instead of $\energy$, as
\begin{equation}\label{Eq:TimePerfectFluidGammaConstant}
t = t_0 + \frac{\kappa_{0}}{\hubble_0} \int_{\deviation_{0}^2}^{\deviation^2} u^{\kappa_1}\,(1-u)^{\kappa_1}\, du \, ,
\end{equation}
where $\kappa_{0}= -2\kappa_{1}\deviation_{0}^{2\kappa_{1}}/3\gamma(1-\deviation_{0}^{2})^{\tfrac{2\kappa_{1}}{\gamma}}$. With the aid of the formula (\ref{Eq:ScaleFactorsExactSolutions}), we obtain the isotropic scale factor,  
\begin{equation}\label{Eq:IsotropicScaleFactorsPerfectFluidGammaConstant}
\frac{a}{a_0} =  \left(\frac{\energy_0}{\energy}\right)^{\frac{1}{3\gamma}} = 
\left(\frac{\deviation_0^2\, (1-\deviation^2)}{\deviation^2\, (1-\deviation_0^2)}
\right)^{\frac{1}{3(2-\gamma)}} \, ,
\end{equation}
and the metric coefficients in terms of $a(t)$, 
\begin{equation}\label{Eq:ScaleFactorsPerfectFluidGammaConstant}
\frac{\ell_k(t)}{\ell_k(t_0)} = C_0\, \left(\frac{a}{a_0}\right) \,  \left(\,
\frac{\sqrt{4\, (a/a_0)^{3(2-\gamma)} + \frac{4\,\deviation_0^2}{1-\deviation_0^2}} + \frac{2\, \deviation_0}{ \sqrt{1-\deviation_0^2}}}{\sqrt{4\, (a/a_0)^{3(2-\gamma)} + \frac{4\,\deviation_0^2}{1-\deviation_0^2}} - \frac{2\, \deviation_0}{ \sqrt{1-\deviation_0^2}}} 
\, \right)^{-\frac{2\, \sin\alpha_k}{3(2-\gamma)}}  \, ,
\end{equation}
where $\alpha_k=\alpha_0+\varphi_k$ and $C_0=[(1-\deviation_0)/(1+\deviation_0)]^{-2\, \sin\alpha_k/3(2-\gamma)}$. If we rename our variables as $a \to R$, $\gamma \to \gamma + 1$, $\alpha \to \psi$ and $\deviation_0 \to \sqrt{\frac{\Omega_0}{1+\Omega_0}}$, %and set 
%\[
%y=\left(\left(\frac{a}{a_0}\right)^{3(2-\gamma)} \,+\, \frac{\deviation_0^2}{1-\deviation_0^2}\right)^{1/2} \, ,
%\]
the formulas (\ref{Eq:TimePerfectFluidGammaConstant}) and (\ref{Eq:ScaleFactorsPerfectFluidGammaConstant}) reproduce  (14a) and (18) in the reference \cite{Jacobs} (see also \cite{ExactSolutions}).
\begin{figure}
\includegraphics[width=0.5\textwidth,height=0.35\textwidth]{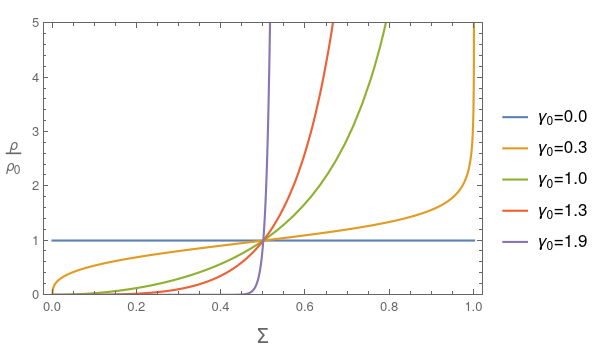}
\hspace{1cm}
\includegraphics[width=0.35\textwidth]{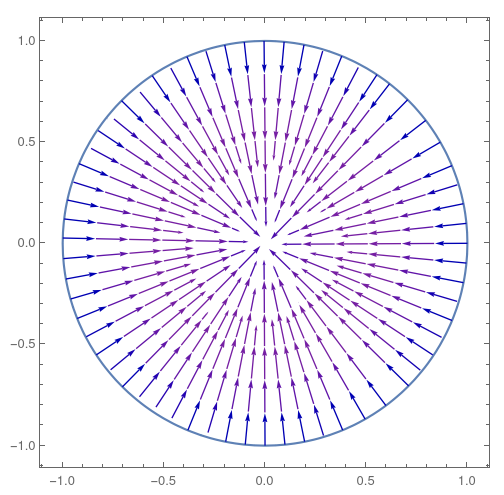}
\caption{On the left, the energy density of the perfect fluid with \\ constant $\gamma$ and $\deviation_0=0.5$. On the right, The Kasner disk for $\gamma=1.3$.}
\label{Fig:PerfectFluidGammaLawPhasePortrait}
\end{figure}

\end{exem}
\begin{exem}\label{Exe:PerfecFluidToy}
Let us consider a toy model with the property that, as the solution approaches the earlier epochs, the matter content becomes stiffer ($\energy\to \infty$ and $\gamma \to 2$), while in the late times it comes closer to the dust model ($\energy\to 0$ and $\gamma \to 1$). A simple way to achieve this is by choosing
\begin{equation}
\pressure = \frac{\energy^2}{\energy + \energy_0}
\quad \textrm{or} \quad
\gamma = \frac{2\, \energy + \energy_0}{\energy + \energy_0} \, .
\end{equation}
\begin{figure}
\includegraphics[width=0.5\textwidth,height=0.35\textwidth]{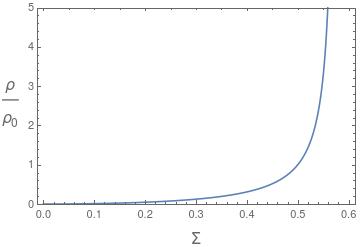}
\hspace{1cm}
\includegraphics[width=0.35\textwidth]{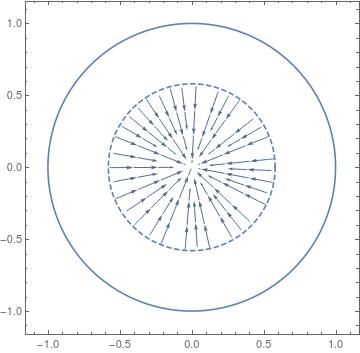}
\caption{On the left, The energy density for the toy model with $\deviation_{0} = 0.5$ \\ and $\deviation_{M} \cong 0.58$. On the right, The Kasner disk for the same model. }
\label{Fig:PerfectFluidToyPhasePortrait}
\end{figure}
In this case, the equation (\ref{Eq:DeviationEnergyRepresentation}) is expressed as
\begin{equation}\label{Eq:DeviationEnergyRepresentationToyModel}
\frac{\energy}{\energy_0} = 
\, \frac{(1-\deviation_{0}^{2})\deviation^{2}}{\, (2+\deviation_{0}^{2})\, (\deviation_{M}^{2} - \deviation^{2})\, }
\quad   
\deviation_M = \deviation_{0}\, \sqrt{\frac{3}{2+\deviation_{0}^{2}}} \, ,
\end{equation}
in the region $0\le \deviation < \deviation_M$.%
\footnote{Note that the solution (\ref{Eq:DeviationEnergyRepresentationToyModel}) in the region $\deviation_M<\deviation$ is spurious, even though $\energy=\energy(\deviation)$ is a well defined function, for the Friedmann equation (\ref{Eq:Friedmann}) would imply that $\hubble^2<0$ there.} The solution becomes singular as $\deviation \to \deviation_M$. This shows that for many solutions there might be a limit anisotropy magnitude $\deviation_M$ which is different from the Kasner's $\deviation=1$. See figure \ref{Fig:PerfectFluidToyPhasePortrait}. 
\end{exem}

\subsection{Extending Kasner: solutions with constant $\Sigma$}\label{Sec:GeneralizedKasner}

Now take $\deviation\ne 0$, $\gamma$ and $\stressparam$ constant. From the equation (\ref{Eq:AnisotropyMagnitudeEquation}), for $0 <\deviation<1$, we get 
\begin{equation}
\stressparam \, \cos\anisophase = (2-\gamma)\, \deviation \, ,
\end{equation}
implying that $\anisophase$ is also constant along any solution. As we use the generalized Friedmann equation (\ref{Eq:Friedmann}) and integrate the equation (\ref{Eq:HubbleDot}) with the suitable choice $t_{-}=0$,  we obtain
\begin{equation}\label{Eq:densityandhubleGeneralizedKasner}
\frac{\energy}{\energy_0}=\left(\frac{\hubble}{\hubble_0}\right)^2= \left(\frac{t_0}{t}\right)^2 
\quad , \quad  \frac{1}{\hubble_0\,t_0}= \frac{3}{2}\,(\gamma+(2-\gamma)\deviation^2)\, .
\end{equation}
Using the equation (\ref{Eq:densityandhubleGeneralizedKasner}) in (\ref{Eq:AngularFrequency}), we get 
\begin{equation}\label{Eq:AngloKasnerGeneral}
\alpha(t) = \alpha_{0} + 
\left(\frac{(1-\deviation^{2})\, \stressparam\,\sin\anisophase}{\deviation\, (\gamma+(2-\gamma)\deviation^2)} \right)\, \ln\left(\frac{t}{t_0}\right)\, .
\end{equation}

In the asymptotic past $t\to 0$ with $0 <\deviation<1$, the solution is clearly bounded away from $\deviation=1$, so that the existence of an oscillatory behavior when $\gamma\ge 0$ and $\stressparam\sin\anisophase\ne 0$ was predicted by the theorem \ref{Thm:PastBehaviourGeneral}, item (iii). Note that $\omega \to \infty$ as $t\to 0$. This can be seen as a consequence of the law $\omega^2 =\kappa \,\energy$, $\kappa$ constant,  which plays the role similar to the conservation of the angular momentum in classical mechanics: as the matter agglomerates, it spins faster. On the other hand, in the asymptotic future $t \to \infty$ with $0 <\deviation<1$,  the solution is clearly bounded away from $\deviation=0$, so that the absence of an oscillatory behavior when $\gamma\ge 0$ was predicted by the theorem \ref{Thm:FutureBehaviourOscillatory}, item (iii). Note also that the formula (\ref{Eq:AngloKasnerGeneral}) shows us that this does not ensure a definite limit for $\alpha$.

\begin{figure}
\includegraphics[width=0.5\textwidth,height=0.35\textwidth]{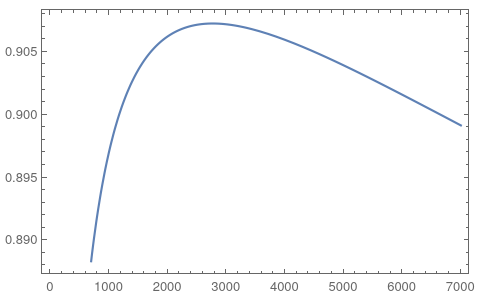}
\hspace{1cm}
\includegraphics[width=0.35\textwidth]{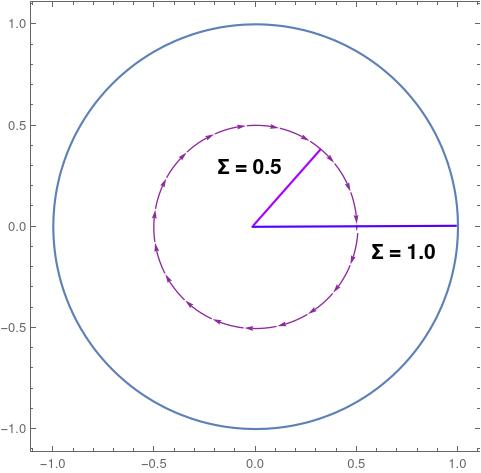}
\caption{Extended Kasner solutions. On the left, $p_1(t)$ for $\gamma=1.8$ , $\stressparam=0.2$\\ and $\deviation_0=0.5$. On the right, the solution representation in the Kasner disk.}
\label{Fig:ExtendedKasner}
\end{figure}
From the formula (\ref{Eq:GeneralMetric}), we can express the general solution with $\stressparam\sin\anisophase\ne 0$ as ($\ell_k(t_0)=t_0^{p_k}$)
\begin{equation}\label{Eq:MetricGeneralizedKasner}
\metric = -\,dt^2 \,+
t^{2\, p_1} dx^2 \,+\, t^{2\, p_2} dy^2 \,+ \, t^{2\, p_3} dz^2
\, ,
\end{equation}
where, after denoting $\varphi_{k+1}(t)=\frac{2k\pi}{3}+\alpha(t)$, 
\begin{equation}
p_{k}(t)=\hubble_0\,t_0  -\, \frac{4 \, \deviation^2}{3\, (1-\deviation^{2})\, \stressparam\,\sin\anisophase} 
\left( 
\frac{\cos\left( \varphi_k(t) \right)- \cos\left( \varphi_k(t_0) \right)}
{\ln\left(t\right)-\ln\left(t_0\right)}
\right)\, .
\end{equation}
We have $p_k(t)\to  \tfrac{1}{3}(1+2\sin(\varphi_k(t_0)))$ in the limit $\deviation_0 \to 1$, which turns the formula (\ref{Eq:MetricGeneralizedKasner}) into the usual Kasner metric \cite{ellis_mac_marteens}. This also happens if $\sin\anisophase=0$.

\subsection{The linear non-oscillatory matter}\label{Sec:AlphaPlusBetaConstant}
Here we consider the spacetimes which can be seen as the asymptotic regime of a large class of physical models, as some classes of solutions containing the magnetic field \cite{calogero,LeBlanc}. They are characterized by the existence of the asymptotic limits of $\pressure/\energy$ and $\pi_k/\energy$. Hence, we assume in this section that
\begin{equation}
\frac{\pressure}{\energy} \,\, \textrm{and} \,\, \frac{\pi_k}{\energy} \,\, \textrm{are constant}
\,\, \left( \gamma, \, \stressparam , \, \phi=\alpha + \anisophase \,\, \textrm{are constant}\right)\, . 
\end{equation}
The dynamical system formed by these conditions and the equations (\ref{Eq:AnisotropyMagnitudeEquation}) and (\ref{Eq:AlphaEquation}) is better understood in terms of the variables $\deviation$ and $\anisophase$. Whenever we refer to the Kasner disk, we readily recover $\alpha$ by $\anisophase=-\alpha+\phi$. 

From the uniqueness property of the solutions with prescribed initial values, we conclude that any solution falls in one of the three classes: $\anisophase=0$, $0<|\anisophase|<\pi$ or $\anisophase=\pi$. An equilibrium point exists only if $0\le \stressparam \le |2-\gamma|$. In such case, it is placed at $\deviation= \deviation_{c}$ and $\alpha=\phi$, if $\gamma <2$, or $\deviation= - \deviation_{c}$ and $\alpha=\phi-\pi$, if $\gamma >2$, where
\begin{equation}
\deviation_{c} = \frac{ \stressparam}{2-\gamma} \, .
\end{equation}
The generic solutions with $\sin\anisophase_0\ne 0$ can be integrated in terms of $\anisophase$, with 
\begin{equation}\label{Eq:AlphaPlusBetaConstantDeviation}
\deviation(t) = \frac{\deviation_c\, \cos\psi_0}{\cos\left(\psi_0- \anisophase(t)\right)} 
\quad , \quad 
\psi_0=\tan^{-1}\left( \frac{\deviation_c - \deviation_0 \cos\anisophase_0}{\deviation_0 \sin\anisophase_0}\right) \, .
\end{equation}
In the limit case $\gamma=2$, $\deviation\sin\anisophase$ is constant. As we define the functions
\begin{equation}\label{eq:1_alpha_beta}
    F_1(u)\,:=\, \int_{\anisophase_0}^u\,  \frac{\cot\anisophase \, d\anisophase}{\cos^2(\psi_0 - \anisophase)-\deviation_c^2\,\cos^2\psi_0} 
\end{equation}
and
\begin{equation}\label{eq:2_alpha_beta}
F_2(u)\,:=\, 
\int_{\anisophase_0}^u\,  \frac{d\anisophase}{\cos^2(\psi_0 - \anisophase)-\deviation_c^2\,\cos^2\psi_0} \, ,
\end{equation}
we can obtain the energy density as
\begin{equation}\label{Eq:EnergyDensityAlphaPlusAnisophaseConst}
    \energy(\anisophase)\,=\,\energy_0 \exp{\left(\, A_1\, F_1(\anisophase)+ A_2\, F_2(\anisophase)\, \right)}
\end{equation}
where, 
\begin{equation}
A_1= \frac{2}{(2-\gamma)^2} \left((2-\gamma)\gamma\,+\, \stressparam^2\right) \cos^2\psi_0
\quad \textrm{and} \quad
A_2= \tfrac{\gamma}{2-\gamma} \sin\, (2\psi_0) \, .
\end{equation}
As we use (\ref{Eq:AlphaPlusBetaConstantDeviation}) in the equation (\ref{Eq:AlphaEquation}) and make the coordinate change $\energy \to \anisophase$ in the integrals (\ref{Eq:TimeDependenceEnergy}) and (\ref{Eq:ScaleFactorsExactSolutions}), we obtain the complete solution as  
\begin{equation}
\frac{C_1}{2}\, (t - t_0) = 
 \int^{\anisophase(t)}_{\anisophase_0} \frac{1}{\sqrt{3 \energy(\anisophase)(\cos^2(\psi_0-\anisophase)-\deviation_c^2\,\cos^2\psi_0)}} \, \frac{d \anisophase}{\sin \anisophase} \, ,
\end{equation}
with $C_1=(2-\gamma)/\cos\psi_0$,  and
\begin{equation}
    \ell_k(\anisophase)\,=\,\ell_k(\anisophase_0) \, \exp{\left(\, B_1\, F_1(\anisophase)+ B_2\, F_2(\anisophase)\, \right)}\, ,
\end{equation}
where 
\begin{equation}
\left\{\begin{array}{lcl}
B_1&=& -\frac{2 \cos^2\psi_0}{3(2-\gamma)^2}\left(2-\gamma + 2\stressparam \sin \left(\phi + \tfrac{2k\pi}{3} \right) \right)
\\[2ex]
B_2&=& \frac{1}{3(2-\gamma)^2}\, \left(4\,\stressparam \cos(\phi + \tfrac{2k\pi}{3})\,\cos^2\psi_0-(2-\gamma) \, \sin(2\psi_0)\right)
\end{array}\right.
\end{equation}

As the numerical parameters are chosen, the functions in (\ref{eq:1_alpha_beta}) and (\ref{eq:2_alpha_beta}) can be easily represented as finite combinations of simpler ones. Nevertheless, as the formulas get more and more complicated, it is questionable if ``solving" the integrals in the generic context is worthwhile. For the sake of comparison, we could have written the energy density in the formula (\ref{Eq:EnergyDensityAlphaPlusAnisophaseConst}) as
\begin{equation}
\begin{array}{lcl}
\energy(\anisophase)&=& c_0\left(\frac{\stressparam+c_2\stressparam\sec \anisophase\,\sin(\anisophase-2\psi_0) - c_1c_2 \tan \anisophase}{\stressparam-c_2\stressparam\sec \anisophase\,\sin(\anisophase-2\psi_0) + c_1c_2 \tan \anisophase}\right)^{c_5 c_3}\, \left(\frac{1-c_4 \tan(\anisophase-\psi_0)}{1+c_1 \tan(\anisophase-\psi_0)}\right)^{c_6} \\[2ex]
&\times& \left((c_1+ \cos(2(\anisophase-\psi_0)))\sin^2 \anisophase \right)^{c_3} \, ,
\end{array}
\end{equation}
where $c_0$ is properly chosen as $\energy(\anisophase_0)=\energy_0$ and
\begin{equation}
\left\{
\begin{array}{ll}
c_1  =1-2\deviation_c^2 \cos^2\psi_0,\quad &
c_2  = \frac{1}{\sqrt{|1-c_1^2|}} \\[2ex]
c_3  =-\frac{(c_1 -1)(\deviation_c + \gamma)}{\deviation_c\,\stressparam\,(c_1+\cos(2\psi_0))} &
c_4  = (c_1-1)c_2, \\[2ex]
c_5  = \sin (2\psi_0) c_2  &
c_6  = \frac{2c_2 \deviation_c \gamma \sin \psi_0}{\stressparam}
\end{array}
\right .
\end{equation}
\begin{figure}
\includegraphics[width=0.35\textwidth]{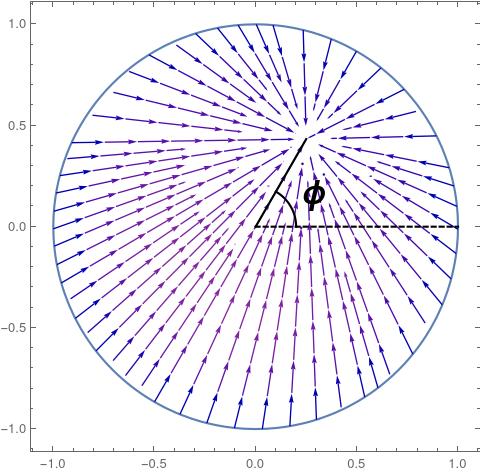}
\hspace{2cm}
\includegraphics[width=0.35\textwidth]{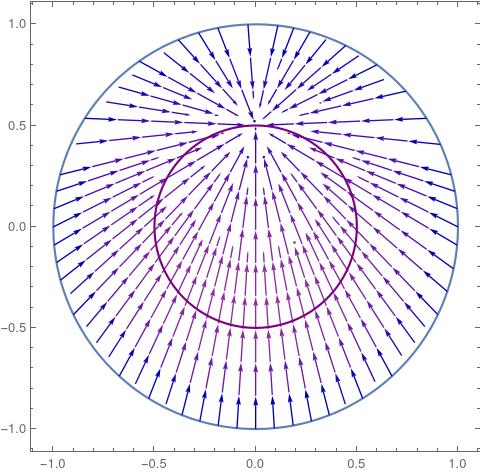}
\caption{For fixed $\gamma$ and $\stressparam$, the equilibrium point rotates with $\phi=\alpha+\anisophase$.}
\end{figure}
Note that the equation (\ref{Eq:AlphaPlusBetaConstantDeviation}) turns out to be affine in the "Cartesian" coordinates $u_1= \deviation\, \cos\alpha$ and $u_2=\deviation\, \sin\alpha$. In other words, the dynamics in the Kasner disk is represented by straight lines. This implies that we have four categories of different qualitative behaviour: $A_{0}$, when $\deviation_{c}=0$, and we return to the perfect fluid case $\stressparam=0$ treated in the example \ref{Exe:PerfectFluidConstantGamma},  $A$, when $0<|\deviation_{c}|<1$, $B$, for $|\deviation_{c}|=1$, and $C$ if $|\deviation_{c}|>1$. In fact, this classification in the literature appears in the cases for which  $\phi$ attains one of the values $\pm \, \pi/6, \pm \, \pi/2, \pm \, 5\pi/6 $, with $A$ divided into $A_{+}$ ($\phi>0$) and $A_{-}$ ($\phi<0$), and so on. The class $C$ is also divided into $1<|\deviation_{c}|<2$ and $|\deviation_{c}|\ge 2$ (D) \cite{calogero}. 
\begin{figure}
\includegraphics[width=0.3\textwidth]{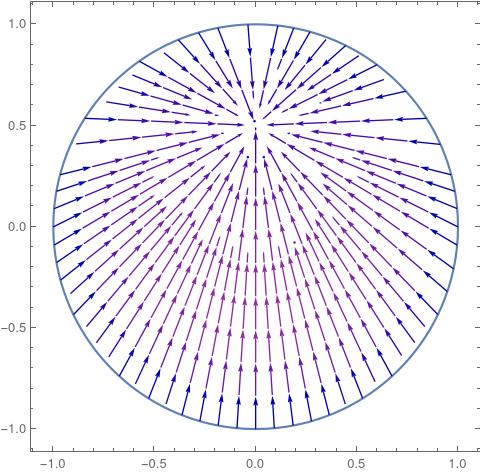}
\includegraphics[width=0.3\textwidth]{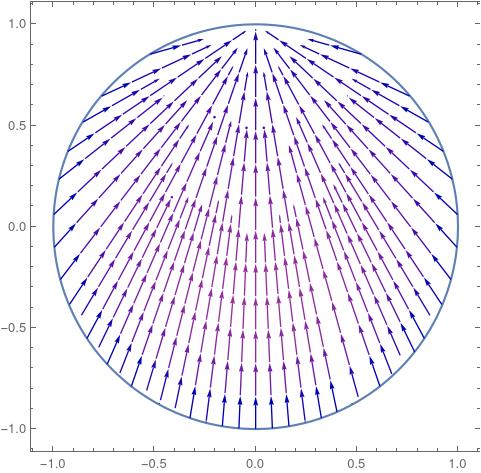}
\includegraphics[width=0.3\textwidth]{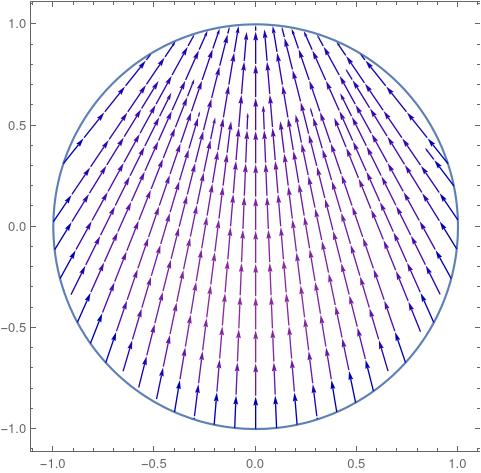}
\caption{The three qualitatively different cases: \\ A ($|\deviation_c|<1$),  B ($|\deviation_c|=1$) and C ($|\deviation_c|>1$).}
\end{figure}

Assuming the expanding dynamics ($\hubble >0$) in the ``$+$" categories, where $\cos\anisophase>0$, we have that $A_{+}$ is characterized by the parameter $\xi_1$ defined in the equation (\ref{Eq:ParameterQualitative}). If $\xi_1 < 2$, then the solution is Kasner in the past and converging to the equilibrium point at $\deviation=\deviation_c$ in the future. If $\xi_1 > 2$, then the solution is Kasner in the future and converging to the equilibrium point at $\deviation=\deviation_c$ in the past. This follows from the equation (\ref{Eq:DeviationDot}), that gives the signal of $\dot{\deviation}$ as we approach the kasner circle. On the other hand, in the categories $B_{+}$ and $C_{+}$, the solutions are Kasner in the past and future. In the special case of the pure magnetic field solution, where $\gamma=\stressparam= 4/3$ and $\phi=\pi/6 (\mod 2\pi/3)$, we obtain $\deviation_c\,=\,2$ (category $D_{+}$ in the reference \cite{calogero}). In this case, by equation (\ref{Eq:AlphaPlusBetaConstantDeviation}),  $|\psi_0|$ is restricted to the interval  $\frac{\pi}{4}<|\psi_0|<\frac{\pi}{2}$. In particular, the integrals (\ref{eq:1_alpha_beta}) and (\ref{eq:2_alpha_beta}) have different functional forms as $\frac{\pi}{4}<|\psi_0|<\frac{\pi}{3}$, $|\psi_0|=\frac{\pi}{3}$ or $\frac{\pi}{3}<|\psi_0|<\frac{\pi}{2}$.  The equation (\ref{Eq:AlphaPlusBetaConstantDeviation}) tells us that as $\deviation(t) \to 1$ we have
\begin{equation}
\anisophase(t) \to \anisophase_1 = \psi_0 \pm \arccos \left(\deviation_c\, \cos\psi_0 \right) \, .
%\quad \Rightarrow \quad \omega = \dot{\anisophase} \to 0\, .
\end{equation}

\subsection{The viscosity as a function of the anisotropy magnitude.}

The formalism of the Newtonian thermodynamics in general relativity \cite{ellis_mac_marteens} demands an equation of state in the form $\pi_{ij} = - \lambda\, \sigma_{ij}$, where $\lambda  \ge 0$ is the viscosity coefficient. Applying it to our variables, we readly conclude that $\anisophase = \pi$ and 
\begin{equation}\label{Eq:ViscosityCoef}
\lambda = \frac{1}{2}\, \frac{\stressparam}{\deviation}\,\sqrt{3\, (1-\deviation^2)\, \energy\,} 
        = \pm \frac{3}{2}\, \, \frac{\stressparam}{\deviation}\,(1-\deviation^2)\, \hubble \qquad (\hubble=\pm|\hubble|)\, .    
\end{equation}
If we allow $\anisophase$ to take other values as we keep the formula (\ref{Eq:ViscosityCoef}) for the viscosity, we arrive to the simple and interesting relation connecting the polar frequency $\omega$, given in (\ref{Eq:AngularFrequency}), with the viscosity, 
\begin{equation}\label{Eq:AngularfrequencyViscosityCoef}
\omega= \pm \, \lambda\, \sin\anisophase \, .    
\end{equation}
Based on these considerations, we would like to set an equation of state in the form $\lambda \sim \hubble$, as would be a natural choice based on their dimensions of frequency and also physically reasonable. For the sake of simplicity, we choose $\gamma$, $\anisophase$ and $\lambda/\hubble$ to be a functions of $\deviation$ only. This is equivalent to
\begin{equation}\label{Eq:EqStateAnisotropyDecoupledViscosity}
\gamma=\gamma(\deviation)\, , \quad  \anisophase=\anisophase(\deviation) 
\quad \textrm{and} \quad \stressparam=\stressparam(\deviation)\, .  
\end{equation}
When $\anisophase=\pi$, we say that we are in the Newtonian limit. 

The dynamical system formed by the equations (\ref{Eq:AnisotropyMagnitudeEquation}) and (\ref{Eq:AlphaEquation}) has the distinguished points determined by the roots of the equation
\begin{equation}\label{Eq:ViscositySolutionRoots}
\stressparam(\deviation) \, \cos\anisophase(\deviation)=(2-\gamma(\deviation))\, \deviation \, .
\end{equation}
The solutions with $\deviation_0$ satisfying the equation  (\ref{Eq:ViscositySolutionRoots}) are all of constant anisotropy magnitude, just like those in the section \ref{Sec:GeneralizedKasner}. In particular, they form a circle of equilibrium points in the cases where $\sin\anisophase(\deviation_0)=0$. If $0<\deviation_0<1$ does not satisfy (\ref{Eq:ViscositySolutionRoots}), then we can integrate the equation $d\deviation/d\alpha$ as
\begin{equation}\label{Eq:AngleViscosityFluid}
\alpha(t) = \alpha_0 - 
\int_{\deviation_0}^{\deviation(t)}\,
\left(\frac{\stressparam\, \sin\anisophase}{(2-\gamma)\, \deviation - \stressparam\, \cos\anisophase}\right)
\, \frac{d\deviation}{\deviation}\, .
\end{equation}
This gives the polar equation of the solution in the Kasner disk. Since the equation (\ref{Eq:AnisotropyMagnitudeEquation}) is separable, we get   
\begin{equation}\label{Eq:EnergyViscosityFluid}
\ln\left(\frac{\energy(t)}{\energy_0}\right) = 2\, \int_{\deviation_0}^{\deviation(t)}\,
\left(\frac{\gamma +  \deviation\,\stressparam\,  \cos\anisophase}{(2-\gamma)\, \deviation - \stressparam\, \cos\anisophase}\right)
\, \frac{d\deviation}{(1-\deviation^2)}\, .
\end{equation}
Hence, we find the general solution in terms of the anisotropy magnitude $\deviation$, as the formulas (\ref{Eq:TimeDependenceEnergy}) is expressed as  
\begin{equation}\label{Eq:TimeViscosityFluid}
%\left(\hubble>0\right)\qquad 
t = t_0 \pm \,
\int^{\deviation(t)}_{\deviation_0}
\frac{2\, d\deviation}{((2-\gamma)\, \deviation - \stressparam\, \cos\anisophase) \sqrt{3\, \energy(\deviation)\, (1-\deviation^2)}}
\end{equation}
and (\ref{Eq:ScaleFactorsExactSolutions}) as
\begin{equation}\label{Eq:ScaleFactorsViscosityFluid}
b_{k+1}(t)=  - \frac{4}{3}\, \int^{\deviation(t)}_{\deviation_0} \,  
\left(
\frac{1+ 2\, \deviation\, \sin\left(\alpha + \num{2k\pi/3} \right)}{(1-\deviation^2)\,((2-\gamma)\, \deviation - \stressparam\, \cos\anisophase)}
\right) 
\, d\deviation \, .
\end{equation}
Therefore, all the exact solutions with $\dot{\deviation}_0\ne 0$ satisfying (\ref{Eq:EqStateAnisotropyDecoupledViscosity}) can be given in terms of $\deviation$ through the formulas from  (\ref{Eq:AngleViscosityFluid}) to (\ref{Eq:ScaleFactorsViscosityFluid}) as
\begin{equation}\label{Eq:MetricViscosityFluid}
\begin{array}{ll}
\metric  &=  - \, \tfrac{4\, d\deviation^2}{3\, (1-\deviation^2)\, ((2-\gamma)\,\deviation - \stressparam\, \cos\anisophase)^2 \energy(\deviation)} + a(\deviation)^2\, d\ell^2\\[2ex]  
d\ell^2&=
e^{2\, G_s(\deviation)}\, dx^2
+ e^{-\, G_s(\deviation) - \sqrt{3}\,G_c(\deviation)}\, dy^2
+ e^{-\, G_s(\deviation) + \sqrt{3}\,G_c(\deviation)}\, dz^2 \, ,
\end{array}
\end{equation}
where the isotropic scale factor is given by
\begin{equation}\label{Eq:expansion_vicosity}
a(\deviation)= \exp{  
\left( - \frac{2}{3}\, \int^{\deviation}_{\deviation_0} \, 
\frac{du}{(1-u^2)\,((2-\gamma(u))\, u - \stressparam(u)\, \cos\anisophase(u))}
\right) }\, ,
\end{equation}
while the remaining functions are
\begin{equation}
G_s(\deviation)= \frac{4}{3}\, \int^{\deviation}_{\deviation_0} \,  
\frac{u\, \sin\left(\alpha(u)\right) \, du}{(1-u^2)\,((2-\gamma(u))\, u - \stressparam(u)\, \cos\anisophase(u))}
\end{equation}
and 
\begin{equation}
G_c(\deviation)= \frac{4}{3}\, \int^{\deviation}_{\deviation_0} \,  
\frac{u\, \cos\left(\alpha(u)\right) \, du}{(1-u^2)\,((2-\gamma(u))\, u - \stressparam(u)\, \cos\anisophase(u))}
 \, .
\end{equation}
For the sake of completeness, let us analyse those exact solutions and their qualitative behaviour as we specialize our equations of state.

\begin{exem}\label{Exem:StressConstant}
Let us assume all the variables in the equation (\ref{Eq:EqStateAnisotropyDecoupledViscosity}) to be constant 
\begin{equation}\label{Eq:EqStateAnisotropyDecoupledViscosityConstant}
\gamma\, ,  \anisophase \, \textrm{and} \, \stressparam \, \textrm{are constant.}
\end{equation}
Since $\stressparam_0=0$ implies the perfect fluid case, dealt with in the section \ref{Sec:PerfectFluid}, we assume $\stressparam\ne 0$. According to the equation (\ref{Eq:ViscositySolutionRoots}), the solution with constant $\deviation(t)=\deviation_0$ and $\gamma_0\ne 2$ will appear if, and only if, $\deviation_0 = \deviation_c$ and $|\deviation_c|\le 1$, where
\begin{equation}\label{Eq:ViscositySolutionRootsExample}
\deviation_c = \frac{\xi-\gamma}{2-\gamma}  \, , 
\quad \xi = \gamma + \stressparam \cos{\anisophase}  
\quad (\gamma\ne 2)\, .
\end{equation}
If $\gamma=2$, this happens only if $\cos\anisophase=0$. In this case, all the solutions have the form $\deviation=\deviation_0$, just like those of the subsection \ref{Sec:GeneralizedKasner}. Hence, we will consider only the cases where $\gamma - 2$ and $\cos\anisophase$ do not vanish simultaneously, so that, as $\deviation\ne \deviation_c$, we obtain from the integral (\ref{Eq:EnergyViscosityFluid}) for $\deviation_c \ne 1$ ($\xi \ne 2$)  
\begin{equation}\label{Eq:EnergyViscosityFluidStressConstant}
\frac{\energy(\deviation)}{\energy_0}\, =  
\left(\frac{1-\deviation}{1-\deviation_0}\right)^{-\, \frac{\xi}{2-\xi}}\,
\left(\frac{1+\deviation}{1+\deviation_0}\right)^{\frac{\xi -
2\gamma}{\xi + 2(1 - \gamma)}}\left( \frac{\deviation_c - \deviation}{\deviation_c - \deviation_0} \right)^{-2\frac{(\xi-\gamma)^2 + \gamma(2-\gamma)}{(\xi-\gamma)^2 - (2-\gamma)^2}}\, .
\end{equation}
The polar equation for the solutions (\ref{Eq:AngleViscosityFluid}) can be written as
\begin{equation}\label{Eq:AngleViscosityFluidNuConstant}
\alpha(\deviation) - \alpha_0\, = 
 \ln\left(\, \frac{\deviation_0\, (\deviation_c - \deviation)}{\deviation\, (\deviation_c - \deviation_0)} \, 
\right)
\, \tan{\anisophase} \, .
\end{equation}
Using these two equations, we readily obtain the isotropic scale factor
\begin{equation}\label{Eq:MetricViscosityFluidStressConstantScaleFactor}
a(\deviation) =  
\left(\frac{1-\deviation}{1-\deviation_0}\right)^{\frac{1}{3(2-\xi)}}
\left(\frac{1+\deviation}{1+\deviation_0}\right)^{\frac{1}{3( 2 +\xi-2\gamma)}}
\left( \frac{\deviation_c - \deviation}{\deviation_c - \deviation_0} \right)^{\frac{2(2-\gamma)}{3((\xi-\gamma)^2-(2-\gamma)^2)}}\, ,
\end{equation}
while the viscosity coefficient turns out to be
\begin{equation}\label{Eq:ViscosityCoefStressConstant}
\frac{\lambda}{\lambda_0}= \, \frac{\deviation_0}{\deviation}
\left(\frac{1-\deviation}{1-\deviation_0}\right)^{\frac{1-\xi}{2-\xi}}
\left(\frac{1+\deviation}{1+\deviation_0}\right)^{\frac{\xi+1-2\gamma}{\xi + 2(1-\gamma)}}
\left( \frac{\deviation_c - \deviation}{\deviation_c - \deviation_0} \right)^{-\frac{(\xi-\gamma)^2+\gamma(2-\gamma)}{(\xi-\gamma)^2-(2-\gamma)^2}}\, . 
\end{equation}
The time component of the metric (\ref{Eq:MetricViscosityFluid}) is 
\begin{equation}\label{Eq:MetricViscosityFluidStressConstant00}
\frac{\metric_{00}}{\metric_{00}(\deviation_0)} =
\left(\frac{1-\deviation}{1-\deviation_0}\right)^{\frac{2(\xi-1)}{2-\xi}}\,
\left(\frac{1+\deviation}{1+\deviation_0}\right)^{\frac{2(2\gamma-1-\xi)}{\xi + 2(1 - \gamma)}}\left( \frac{\deviation_c - \deviation}{\deviation_c - \deviation_0} \right)^{\frac{4(2-\gamma)}{(\xi-\gamma)^2 - (2-\gamma)^2}}\, .
\end{equation}
The solution is completed by the functions $G_s(\deviation)$ and $G_c(\deviation)$. They attain the simplest form as we keep their integral representations, rather than trying to resolve them in terms of elementary functions: 
\begin{equation}
G_s(\deviation)= \frac{4}{3(2-\gamma)}\, \int^{\deviation}_{\deviation_0} \,  
\frac{u\, \sin\left(
\alpha_0\, +
 \ln\left(\, \frac{\deviation_0\, (\deviation_c - u)}{u\, (\deviation_c - \deviation_0)} \, 
\right)
\, \tan{\anisophase} \,
\right) \, du}{(1-u^2)\,( u - \deviation_c)} \, .
\end{equation}
The $G_c(\deviation)$ has an analogous formula, just with the sine exchanged by the cosine. 
\begin{figure}
\begin{center}
\includegraphics[width=0.35\textwidth]{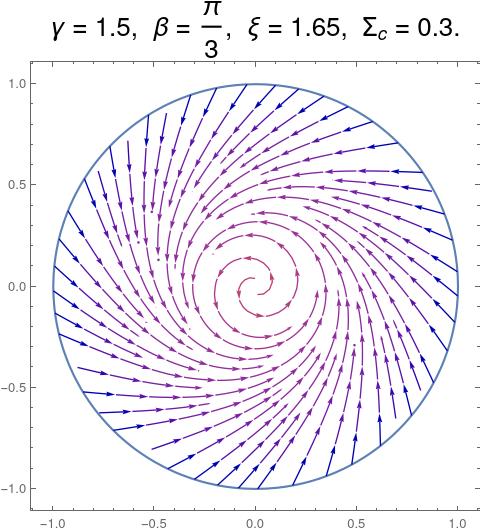}\hspace{2cm}
\includegraphics[width=0.35\textwidth]{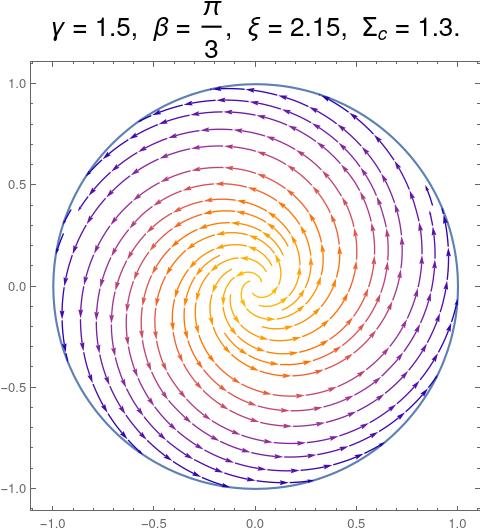}\\
\includegraphics[width=0.35\textwidth]{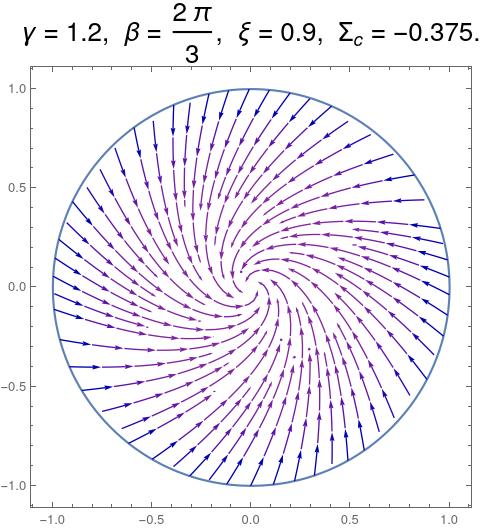}\hspace{2cm}\includegraphics[width=0.35\textwidth]{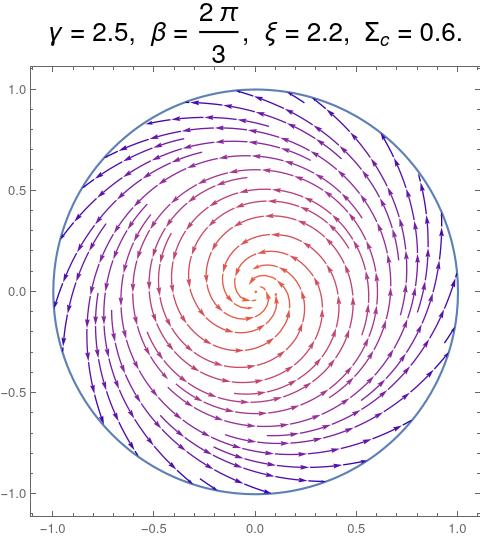}%\end{array}
\end{center}
\caption{Qualitative behaviour for $\gamma$, $\stressparam$ and $\anisophase$ constant.}
\label{Fig:GamaStressAnisophaseConstant}
\end{figure}

There are three distinct regions to be analysed in the asymptotic regime with $\hubble>0$:
\begin{itemize}
\item {\bf Kasner region}: According to (\ref{Eq:MetricViscosityFluidStressConstantScaleFactor}),  $a \to 0$ if $\xi<2$ and $a \to \infty$ if $\xi>2$. Therefore, we are in the asymptotic past when $\xi<2$ and in the asymptotic future when $\xi>2$. From the equations (\ref{Eq:AngularfrequencyViscosityCoef}), (\ref{Eq:EnergyViscosityFluidStressConstant}) and (\ref{Eq:ViscosityCoefStressConstant}),  as we approach the Kasner circle $\deviation=1$, we have the following:   
\begin{enumerate}[(a)]
\item \underline{$\deviation \to 1$ with $\xi<1$}: Asymptotic past ($a \to 0$) with no oscillation ($\lambda \to 0$). This illustrates the first item in the theorem \ref{Thm:PastBehaviourGeneral}. Note that for $\xi<0$ we have the non-physical condition $\energy\to 0$, which is a consequence of $\dot{\energy}\hubble >0$. See the third picture in the figure \ref{Fig:GamaStressAnisophaseConstant}.
\item \underline{$\deviation \to 1$ with $1<\xi<2$}: Asymptotic past ($a \to 0$) with the weak mixmaster behaviour for $\sin\anisophase \ne 0$($\lambda \to \infty$). This represents an anisotropic primordial universe ($\energy\to \infty$), and illustrates the second item in the theorem \ref{Thm:PastBehaviourGeneral}. See the first picture in the figure \ref{Fig:GamaStressAnisophaseConstant}.  
\item \underline{$\deviation \to 1$ with $\xi>2$}: Asymptotic future ($a \to \infty$) with no oscillatory behaviour  ($\lambda \to 0$). This represents an anisotropic late time universe ($\energy\to 0$), which illustrates the last item in the theorem \ref{Thm:FutureBehaviourOscillatory} in the case  $\gamma \ge 0$. See the second and fourth pictures in the figure \ref{Fig:GamaStressAnisophaseConstant}.
\end{enumerate}
\item {\bf Isotropic region}: $\deviation=0$ is an equilibrium point of the system, which we readily see by inspecting the equations for $\dot{x}$ and $\dot{y}$, with $x=\deviation\cos\anisophase$ and $y=\deviation\sin\anisophase$ their Cartesian representation. According to the equations (\ref{Eq:EnergyViscosityFluidStressConstant}) and (\ref{Eq:MetricViscosityFluidStressConstantScaleFactor}),  $\energy$ and $a$ tend to a finite value as $\deviation \to 0$.  By a direct inspection of the formula (\ref{Eq:expansion_vicosity}),
\begin{equation}
 \left(\frac{1}{a}\frac{da}{d\deviation}\right)_{\deviation \to 0} = \frac{2}{3 \stressparam \cos \anisophase} \, .   
\end{equation}
Since $\hubble>0$, we conclude that $\deviation=0$ represents the asymptotic past when $\cos\anisophase > 0$ and the asymptotic future when $\cos\anisophase < 0$. From the equations (\ref{Eq:AngularfrequencyViscosityCoef}) and (\ref{Eq:ViscosityCoefStressConstant}), we have that the solution is pulsating if $\sin\anisophase \ne 0$ ($\omega \to \infty$). Hence, we have the following:   
\begin{enumerate}[(a)]
\item \underline{$\deviation \to 0$ with $-1< \cos\anisophase < 0$}: Asymptotic future with an oscillatory behaviour. This represents an isotropic late time universe with $\energy\to A\, \energy_0$, $A\ne 0$. It illustrates the first item in the theorem \ref{Thm:FutureBehaviourOscillatory}. See the fourth picture in the figure \ref{Fig:GamaStressAnisophaseConstant}.
\item \underline{$\deviation \to 0$ with $0< \cos\anisophase < 1$}: The solution is pulsating in the past with $\omega \to \infty$. This represents an isotropic primordial universe with both the scale factor and the energy density bounded, and illustrates the third item in the theorem \ref{Thm:PastBehaviourGeneral}. See the first and second pictures in the figure \ref{Fig:GamaStressAnisophaseConstant}.
\end{enumerate}
\item {\bf $\deviation_c$ region}: Here we consider only the case $0<\deviation_c<1$. We have the three distinct cases, where we define  $\gamma_{-}=1-\sqrt{1+\stressparam^2\cos^2\anisophase}$:  
\begin{enumerate}[(a)]
\item \underline{$\deviation \to \deviation_c$ with $\gamma<\gamma_{-}<0$}: this is the non-physical case, for $\dot{\energy}\, \hubble>0$ along the solution. This follows from the fact that $\gamma + \deviation_c\, \stressparam\cos\anisophase < 0$. Here we have $\cos\anisophase > 0$, $\xi <2$ and, according to (\ref{Eq:MetricViscosityFluidStressConstantScaleFactor}),  $a \to \infty$, that is, the solutions arrive at $\deviation=\deviation_c$ in the asymptotic future. But $\energy \to \infty$ and, if $\sin\anisophase\ne 0$, $\omega \to \infty$ as well, showing us that $\gamma \ge 0$ is an essential condition in the last statement of the theorem \ref{Thm:FutureBehaviourOscillatory}.  
\item \underline{$\deviation \to \deviation_c$ with $\gamma_{-}< \gamma < 2$}: here we have $\cos\anisophase > 0$, $\xi <2$ and, according to (\ref{Eq:MetricViscosityFluidStressConstantScaleFactor}),  $a \to \infty$. This means that the solutions arrive at $\deviation=\deviation_c$ in the asymptotic future as they come from the Kasner circle ($\deviation_0 > \deviation_c$) or the isotropic point ($\deviation_0 < \deviation_c$) in the asymptotic past. In both cases, the solution is non-oscillatory in the future. This illustrates the last item in the theorem \ref{Thm:FutureBehaviourOscillatory}. See the first picture in the figure \ref{Fig:GamaStressAnisophaseConstant}.
\item \underline{$\deviation \to \deviation_c$ with $\gamma>2$}: here we have $\cos\anisophase < 0$, $\xi >2$ and, according to (\ref{Eq:MetricViscosityFluidStressConstantScaleFactor}),  $a \to 0$. This means that the solutions are at $\deviation=\deviation_c$ in the asymptotic past and move towards the Kasner circle ($\deviation_0 > \deviation_c$) or the isotropic point ($\deviation_0 < \deviation_c$) in the asymptotic future. From (\ref{Eq:ViscosityCoefStressConstant}) and (\ref{Eq:EnergyViscosityFluidStressConstant}), we have $\energy \to \infty$ and, if $\sin\anisophase \ne 0$, the solution is oscillatory in the past with $\omega \to \infty$. This illustrates the last statement in the theorem \ref{Thm:PastBehaviourGeneral}. See the fourth picture in the figure \ref{Fig:GamaStressAnisophaseConstant}.
\end{enumerate}
\end{itemize}
\end{exem}
\section{Final remarks: some physical and conceptual considerations}\label{Sec:PhysicalInterpretation}

In this manuscript, we have focused on the conceptual development concerning the cosmological anisotropy in its simplest and purest form, allowing the space sections to be homogeneous and flat. The starting point was the use of observational ($\hubble$, $\deviation$ and $\alpha$) and physical ($\energy$, $\gamma$, $\stressparam$ and $\anisophase$) variables only. Under this parameters, we have shown that the Einstein equations (\ref{Eq:Friedmann}-\ref{Eq:AlphaEquation}) turn out to be represented in a form much simpler than the one we usually find in the literature (compare with \cite{calogero,LeBlanc}). This is due to the use of the variables $\alpha$ and $\anisophase$, first appeared in \cite{BGK}. Hence, we have taken advantage of this fact to better understand some properties of the most striking parameter of the anisotropy: the angular frequency $\omega$. In particular, as a byproduct, we have developed a large scheme of obtaining exact solutions (section \ref{Sec:ExactSolution}).         

In this final section, we turn to the physical implications of our results. In order to do so, let us fix our physical context throughout as
\begin{equation}
\energy > 0 \, , \quad \hubble > 0 \, , \quad \dot{\energy}<0 \, \quad (\gamma + \deviation\, \stressparam\, \cos\anisophase >0)\, ,
\quad 0<\gamma < 2 \, .
\end{equation}
The usual energy inequalities, as well as some sufficient conditions on $\gamma$ and $\stressparam$ to attain them, is displayed in the figure \ref{Tab:Energyconditions}, for convenience.  
\begin{table}[h]
    \centering
    \begin{tabular}{l||c|c}\label{Tab:Energyconditions}
                 &  Canonical relations & Sufficient conditions \\
\hline
\hline

        Null     &  $\energy + p_{i} \geq 0$ & $\gamma \geq \stressparam$ \quad $\energy \geq 0$ \\
\hline
        Week     &  $\energy + p_{i} \geq 0$ & $\gamma \geq \stressparam$ \\
                &  $\energy \geq 0$ & $\energy \geq 0$ \\
\hline
        Strong   &  $\energy + p_{i} \geq 0$ & $\gamma \geq \stressparam$ \quad $\energy \geq 0$\\
                 &  $\energy + 3\pressure \geq 0$ & $\gamma \geq \tfrac{2}{3}$ \\
\hline
        Dominant &  $\energy \geq |p_{i}| $ & $ 2 \geq \gamma \geq \stressparam$ \quad $\energy \geq 0$
    \end{tabular}\\[2mm]
\caption{
Sufficient inequalities for the energy conditions.}
\end{table}
\subsection{Isotropization and anisotropization}
\label{Sec:IsotropizationAnisotropization}

The usual behaviour attributed to the anisotropy, which is often tacitly assumed to be the case, is that of our example \ref{Exe:PerfectFluidConstantGamma}: if it existed, it would be large ($\deviation \to 1$) close to the initial singularity, and fading away as the time passes by, typically characterized by a term $a^{-6}$ in the Friedmann equation \cite{wainwright}, with no oscillation at all.  In fact, this seems to happen due to the preference of dealing with the perfect fluids in the literature ($\omega=0$). Nevertheless, if the equation of state do not obey a ``$\gamma$-law", this picture cannot be sustained in general (see the example \ref{Exe:PerfecFluidToy}). Moreover, even with a small contribution of the stress $\pi_k\ne 0$ to the ``$\gamma$-law" perfect fluid will usually lead to very different qualitative and quantitative behaviours. %

The isotropization and the anisotropization processes have been known, at least tacitly, since the first considerations of the influence of the magnetic field in the cosmic expansion \cite{Doroc,Vajk,Reinhardt,LeBlanc}. A good example is given in the section \ref{Sec:AlphaPlusBetaConstant}, where, except for the $A_0$ class, the solutions are Kasner in the past, evolve to a more isotropic one until it reaches its minimum anisotropy, at the point in the Kasner disk closest to the origin, and then begin an anisotropization epoch, converging towards the equilibrium point with $\deviation_c \ne 0$  ($\deviation_c > 1$ in the C,D-classes, which means that they are Kasner in the future in those cases). This kind of procedure can be easily understood as we define, along a given solution, the function 
\begin{equation}
\deviation_c(t):= \frac{\stressparam(t)\cos\anisophase(t)}{2-\gamma(t)} \, .
\end{equation}
Therefore, according to the equation (\ref{Eq:AnisotropyMagnitudeEquation}), we have 
\begin{equation}
\deviation(t) > \deviation_c(t) \quad \Rightarrow \quad \textrm{Isotropization epoch,}    
\end{equation}
while 
\begin{equation}
\deviation(t) < \deviation_c(t) \quad \Rightarrow \quad \textrm{Anisotropization epoch.}    
\end{equation}
A good example of this behaviour is represented in the first picture of the figure \ref{Fig:GamaStressAnisophaseConstant}, where $\deviation_c=0.3$. Note that the anisotropization epoch depend much more on how the anisotropies of the spacetime and the fluid are coupled, represented by the value of $\anisophase$, than to their magnitudes $\sqrt{\sigma_{\mu\nu}\sigma^{\mu\nu}}$ and $\sqrt{\pi_{\mu\nu}\pi^{\mu\nu}}$. Furthermore, it can occur with no violation of any one of the usual energy conditions (figure \ref{Tab:Energyconditions}) and even with $\gamma>1$, that is, with no exotic matter at all. 

Only as we fully understand the physical mechanisms responsible for the anisotropization process, we will be in a better position to understand if the tiny values of the anisotropy in the CMB sky \cite{IstSaadeh,Tedesco2019} at redshift $z\sim 1100$ imply an irrelevant contribution of $\deviation$ in the late-time universe at $z \sim 1$. Although this has been tacitly assumed, there is no convincing justification on why considering non-perfect fluids could not be the case, and therefore why an anisotropization epoch cannot occur between the last scattering and the current epoch, or even in other periods. In times when the foundations of the standard cosmological model have been shaken \cite{Riess}, bring back some neglected observational parameters and re-discuss its influence under a new perspective is a scientific duty. Some effort to link the role of $\deviation$ with the observations in the supernovae surveys have appeared recently \cite{Leandro}, and the influence of $\omega$ on the cosmographic parameters is under investigation.

\subsection{The anisotropy converted into energy density}
\label{Sec:EnergyAnisotropyExchange}

The usual anisotropy pattern associated to the $\Lambda$CDM model appears as a $\Omega_\sigma/a^6$ term in the Friedmann equation \cite{Tedesco2019}, $\Omega_\sigma\ge 0$, which is nothing more than the term $3\, \deviation^2\, \hubble^2$ in the equation (\ref{Eq:Friedmann}) for the perfect fluid case. According to the formulas (\ref{Eq:DeviationDot}) and (\ref{Eq:HubbleDot}), it satisfies the equation 
\begin{equation}
 \frac{d\Omega_\sigma}{da} = \frac{3\, (1-\deviation^2)}{a\, \deviation}\, \Omega_\sigma\, \stressparam\, \cos\anisophase \, .
\end{equation}
Hence, we note that for a perfect fluid, $\Omega_\sigma$ is constant, as it is usually assumed to be the case. In this context, nothing interesting seems to appear with some cosmological relevance. But under the non-perfect fluid hypothesis, there are much more remarkable features, for if $\stressparam\cos\anisophase \ne 0$, there is a conversion taking place among the anisotropy magnitude and the energy density, which seems to have been neglected in the literature, except for some punctual examples (see \cite{BGK}, for instance). 

Let us split the energy density in its isotropic part, $\mu$, with $\dot{\mu} = -\, 3\, \hubble \, \mu \, \gamma$, and the internal contribution $\epsilon_\sigma$, due to the conversion of anisotropy into energy. We have, following the conservation equation (\ref{Eq:Conservacao}),  
\begin{equation}\label{Eq:ConservacaoSplit}
\energy = (1+\epsilon_\sigma)\, \mu \, , \quad 
\frac{d\epsilon_\sigma}{da} = -\, 3\, \left(\frac{1+\epsilon_\sigma}{a}\right) \, \deviation \, \stressparam \cos{\anisophase}  \, .
\end{equation}
Again, the parameter $\anisophase$ seems to be some kind of measurement of the coupling between the anisotropies in the spacetime and in the cosmological fluid. It gives us the following interpretation: 
\begin{itemize}
\item \underline{$\cos\anisophase < 0$}: the anisotropy term in the $\Lambda$CDM model decays faster than $1/a^6$. On the other hand, $\epsilon_\sigma$ increases, so that the anisotropy is transformed into energy. Therefore, this leads to a framework where the universe becomes isotropic faster than the perfect fluid case, but where the anisotropy in the past have been transformed into energy, so that its contribution can be perceptible even in the late-time cosmology, at least for some cases. 

In the example \ref{Exem:StressConstant}, where $\gamma$, $\stressparam$ and $\anisophase$ are constant, a straightforward inspection of the formula (\ref{Eq:EnergyViscosityFluidStressConstant}) tells us that $\energy \to \energy_\infty \ne 0$ as $\deviation \to 0$. In this case, $\deviation_c$ is constant and the solution is isotropic in the future with a constant energy density, mimicking the cosmological constant effect in the flat FLRW spacetimes. It is interesting to note that no violation of the usual energy conditions (figure \ref{Tab:Energyconditions}) nor even exotic matter is necessary for this effect. Indeed, this is a common kind of behaviour of the late-time anisotropy (see also \cite{BGK}).  
\item \underline{$\cos\anisophase > 0$}: the anisotropy term in the $\Lambda$CDM model decays slower than $1/a^6$, while $\epsilon_\sigma$ decreases, so that the energy is transformed into anisotropy. It could happen, for instance, along a period of anisotropization. In this case, not only the role of the anisotropy in the $\Lambda$CDM model should be revised, but the entire model itself.  
\end{itemize}

\subsection{A criticism on the definitions for the oscillatory and mixmaster behaviours}\label{Sec:CriticismOscillations}

In this manuscript we have dealt with two concepts that are too evasive to be easily tamed by a precise mathematical definition: the oscillatory and the mixmaster asymptotic behaviours. Hence, it is our duty to expose what we have left out and what seems to be unfit to our expectations. 

The first observation concerns the definition of oscillatory behaviour presented in the section \ref{Sec:AnisotropyTheorems}. The reader might have noted that, in the section \ref{Sec:GeneralizedKasner}, the solutions with constant anisotropy magnitude do not oscillate in the future, according to our definition. This fact occurs because $\alpha - \alpha_0 \sim \ln t$ with $t_{+}=\infty$, which implies $\omega \sim 1/t \to 0$. Indeed, this represents a kind of oscillatory behaviour which is not included in our formulation. However, if we think of the cosmological observations, that are collected at periods where the energy scale does not change\footnote{In the late-time supernovae observations we expect $\energy\sim \energy_0$, while observing the CMB pattern we assume $\energy\sim \energy_L$ (last scattering).},  those solutions, where $\alpha - \alpha_0 \sim \ln \energy$, would hardly be distinguished from $\alpha = \alpha_0$ in an observable basis. 

Another important remark is that the drawings of the phase portraits can be very misleading, as they concern the asymptotic oscillations, for what we see in the pictures is $d\alpha/d\deviation=\dot{\alpha}/\dot{\deviation}$, and not $\omega=\dot{\alpha}$. In principle, it is not possible to determine whether or not the asymptotic oscillation is occurring just by a visual inspection of the phase portrait. The text is full of these misleading pictures.            

The most evasive of the intuitive concepts we are dealing with is the mixmaster one. Whatever its definition might be, it must include any solution having ``alternating Kasner epochs"\cite{Misner,BKL}. Under this guiding situation, we have formulated our concept in the section \ref{Sec:AnisotropyTheorems}. It encompass the chaotic ``billiard" solutions \cite{LeBlanc,BelinskiLivro}, as it should, but also solutions as the one represented in the first picture of the figure \ref{Fig:GamaStressAnisophaseConstant}, which hardly resembles our initial intuition. That is the reason why we have opted for the adjective ``weak" in our definition. The ``true" mixmaster behaviour happens when the kasner circle behaves as a limit circle of the dynamics, so that $\alpha(t)$ does not attain a well defined limit as we approach the singularity. A simple way to throw out many undesired ``mixmaster" solutions is by defining the angle $\mixangle_1$ between the tangent of the trajectory and the ``radial'' vector as the solution approaches the Kasner circle, that is,
\begin{equation}\label{Eq:MixAngle}
\tan \mixangle_1 = \lim_{\deviation\to 1}\, \left|\frac{\dot{\alpha}}{\dot{\deviation}}\right| 
= \lim_{\deviation\to 1}\, \left| \frac{\stressparam\, \sin\anisophase}{\deviation\, (\,(2-\gamma)\, \deviation - \stressparam\, \cos\anisophase)} \right|\, .   
\end{equation}
This would define a ``degree of alternating Kasner epochs": the more $\mixangle_1$ approaches $\pi/2$, the more the solution looks like a ``true" mixmaster one. The class of the intuitive mixmaster examples, whatever this might be, should be close to that of the solutions where $\mixangle_1$ is equal to $\pi/2$ or is not well defined.

\section*{Acknowledgments}

G. A. S. S. and W. O. C. E. acknowledges CAPES/Brazil for the financial support under the grants $\# 88882.441458/2019-01$ and $\# 88882.430236/2019-01$, respectively, in partial fulfillment of the requirements for the degree of Master in Science. B. B. B. is likewise grateful to PIBIC/UNIFEI for the received financial support. 

\end{document}